\documentclass[12pt, letter]{article}
\usepackage[utf8]{inputenc}

\usepackage{comment}
\usepackage{graphicx}
\usepackage{graphics}
\usepackage{hyperref}
\usepackage{amssymb}
\usepackage{amsmath}
\usepackage{slashed}
\usepackage{authblk}
\usepackage{multirow}
\usepackage{float}

\usepackage{subfigure}
\usepackage[usenames,dvipsnames]{color} 
\newcommand\Tstrut{\rule{0pt}{2.6ex}}         
\newcommand\Bstrut{\rule[-0.9ex]{0pt}{0pt}}   
\newcommand{\atau}{$\rm{a_{\tau}}~$}

\begin{document}

\title{\bf A study of the measurement of the  $\tau$ lepton anomalous magnetic moment in high energy lead-lead collisions at LHC}

\author[1]{Monica Verducci,}
\author[2,3]{Natascia Vignaroli,}
\author[4]{Chiara Roda,}
\author[5]{Vincenzo Cavasinni.}

\renewcommand\Affilfont{\fontsize{11}{11}\itshape}

\affil[1,4,5]{Dipartimento di Fisica, Universit\`{a} di Pisa and 
Istituto Nazionale di Fisica Nucleare, Sezione di Pisa, I-56127 Pisa, Italy}

\affil[2]{Dipartimento di Fisica, Universit\`{a} di Napoli and 
Istituto Nazionale di Fisica Nucleare, Sezione di Napoli, I-80126 Napoli, Italy}
\affil[3]{Dipartimento di Matematica e Fisica, Universit\`{a} del Salento, and 
Istituto Nazionale di Fisica Nucleare, Sezione di Lecce, I-73100 Lecce, Italy}

\date{}
\maketitle

\begin{abstract}
The $\tau$ lepton anomalous magnetic moment: $a_\tau = \frac{g_{\tau}-2}{2}$ was measured, so far, with a precision of only several percents despite its highly sensitivity to physics beyond the Standard Model such as compositeness or Supersymmetry. A new study is presented to improve the sensitivity of the $a_\tau $ measurement with photon-photon interactions from ultra-peripheral lead-lead collisions at LHC. The theoretical approach used  in this work is based on an effective Lagrangian and on  a photon flux implemented in the MadGraph5 Monte Carlo simulation. 
Using  a multivariate analysis to discriminate the signal from the background processes, a sensitivity to the anomalous magnetic moment  $\rm{a_{\tau}}$ = 0 $_{+0.011} ^{-0.019}$ is obtained at 95\% CL with a dataset corresponding to an integrated luminosity of 2 nb$^{-1}$ of lead-lead collisions and 
assuming  a conservative 10\% systematic uncertainty. The present results are compared with previous calculations and available measurements.
\end{abstract}

\newpage

\section{Introduction}

The anomalous magnetic moment of elementary particles (leptons and quarks) is defined as: $a_{l,q} = \frac{g_{l,q}-2}{2}$ where, Dirac theory of the QED implies at the classical level, $g_{l,q}=2$. The measurement of  $a_{l,q}$ is today  one of most powerful  tool to test the validity of the the Standard Model (SM) theory that, despite its indisputable success, can not be a complete theory. There are, in fact, several unresolved questions not addressed by the SM
such as, for example, the existence of “dark matter”, estimated to be about 5
times more abundant than the ordinary matter. Precise measurements of the elementary particles magnetic moment and its comparison with the Standard Model predictions could indicate the existence of new interactions and particles  that could shed light also on the nature of the dark matter and on the problems of the naturalness and fine tuning of the Higgs boson mass.  
\par
Extensive researches of new physics, new particles, deviations from the SM predictions, have been carried out at the LHC, but no clear hints of the existence of new phenomena have   emerged  from the data collected so far. The LHC run has re-started in 2022  with the LHC performances improved both in energy and luminosity and searches of new particles  will continue, but the limited  energy/mass  at which they could be produced and detected   will remain  to be of the order of 1 TeV.
\par
A discrepancy between the values of $a_{l,q}$ predicted by the SM and the measured ones could provide important clues to anticipate both the nature and the mass of the new phenomena suggesting also the energy regime at which a direct production of the new particles responsible of these discrepancies could be expected.  This, of course, would be possible, if, in one hand, a precise calculation is provided of the SM predictions, and, on the other hand, accurate measurements of the the anomalous $a_{l,q}$  will be performed for the three charged leptons. 
\par
For an accurate prediction of $a_{l,q}$ within the SM it is crucial a precise evaluation of high order electromagnetic (QED), weak and strong (QCD) interactions  corrections. These QED corrections were first calculated, for the electron, in the seminal  paper by J. Schwinger \cite{Schwinger}  to  be $a_e=\frac{\alpha}{2 \pi}=0.001162$, where $\alpha$  is the fine structure constant.\par The QCD corrections are difficult to calculate in an energy range  where no perturbation development is applicable and corrections should rely on experimental cross sections in  lepton-hadron and $e^{+}e^{-}$ hadronic interactions with the help of the dispersion relation techniques. The size of the hadronic corrections strongly depends on the mass of the lepton under study becoming more and more relevant as the the lepton mass increases.
\par 
The experimental technique to measure $a_{l}$ :  $a_e,a_\mu,a_\tau$, is different for the three charged leptons. An extreme  precision of  $0.28$  ppb for $a_e$ is obtained by a single-electron quantum cyclotron frequency measurements~\cite{Hanneke}. A recent improved observation  of the fine structure constant $\alpha$ led  to a difference between the measured and predicted $a_e$ negative and significant at 2.4 $\sigma$'s ~\cite{Hanneke}.
\par
For the muon, $a_{\mu}$ is measured by comparing the cyclotron  and  the muon magnetic moment precession frequencies.  The recent experiment at Fermilab,   `"Muon g-2", measured $a_\mu$ with a precision of 0.46~ppm \cite{Abi}.  The difference between the SM prediction  of $a_\mu$, with hadronic contributions calculated via the dispersion relation method~\cite{Aoyama}, and the combined``Muon g-2" and E821 at BNL experiments,  shows a discrepancy of 4.2 $\sigma 's$. However, a new estimate of the theoretical predicted value of $a_{\mu}$  obtained by recalculating the hadronic contributions using a lattice  QCD approach, resulted into a theoretical prediction compatible with the experimental value within 1.2 $\sigma's$ \cite{Borsanyi}.
\par The theoretical prediction of $a_{\tau}$, although not as precise as those ones for lighter leptons, is by far more accurate than the experimental measurements. Theoretically, the larger $\tau-$mass makes the hadronic contributions much larger than in  case of the  electron and the muon and, consequently, also the uncertainties of the $a_\tau $ is much larger. Possible contributions to  $a_{l,q}$ given from new particles of mass $M$ to the photon-lepton vertex are expected to be of the order of $m_l^2/M^2$ for a lepton of mass $m_l$. Therefore, new physics effects for the $\tau$ would be enhanced by a factor $m_{\tau}^2/m_{\mu}^2$ = 286 with respect to $\mu$. 
Moreover, in some models addressing recent anomalies in $R_{D^{(*)}}$ \cite{LHCb:2015gmp}, a significant contribution to $a_{\tau}$  could arise from  new scalar and tensor operators and $\Delta a_{\tau}$ could be as large as $10^{-3}$ \cite{Feruglio:2018fxo}.\par The  $a_\tau$  would  also be sensitive to possible lepton compositeness that, in  general, would contribute with  corrections $O(m^2_{\tau}/\Lambda^2_c)$, where $\Lambda_c$ is the "compositeness" scale \cite{Silverman}, possibly generated by warped extra-dimensions \cite{Csaki:2008qq,Chen:2009gy,Kadosh:2010rm, delAguila:2010es, delAguila:1991rm}. 
\par The $a_\tau$ investigation, definitely,  represents an excellent tool to access new physics beyond the SM (BSM).
\par
Unfortunately the present  experimental knowledge of  $a_\tau$ is  poor. In fact the very short $\tau$-lifetime precludes the use of the precession frequency measurement method  as done in the $\mu-$case. The method adopted is to exploit the sensitivity to $a_{\tau}$ of the $\tau-pair$ total and differential cross-sections, in photon-photon scattering\footnote{The first idea of using photons accompanying fast moving charged particles  for physics experiments is due to Enrico Fermi in 1924 \cite{Fermi}. }. 
The best $a_\tau$ measurement at LEP was obtained by the DELPHI experiment \cite{DELPHI:2003nah} and provided the limit $ -0.052 < a_{\tau}^{exp} < 0.013$ at $95\%$ CL.
The combined reanalysis of various experimental measurements such as the $e^+e^-\rightarrow \tau^+ \tau^-$ cross section, the transverse $\tau^-$ polarization and asymmetry, as well
as the decay width $\Gamma (W \rightarrow  \tau^+ \nu_{\tau} )$, allowed the authors in \cite{Gonzales} to set a stronger, but indirect,
model-independent limits on new physics contributions to $a_{\tau}$: $-0.007 <a_{\tau} < 0.005$.  Other strategies to measure the $\tau$ anomalous moment at the LHC have been proposed in \cite{Galon:2016ngp}, by considering the rare Higgs decay $h \to \tau \tau \gamma$, which shows a sensitivity at the percent level and the measurement of the  distribution of the large transverse mass of $\tau-pairs$ produced in proton-proton collisions  \cite{Haisch}.  
\par
\begin{figure}[ht!]
    \centering
   \includegraphics[width=0.5\linewidth]{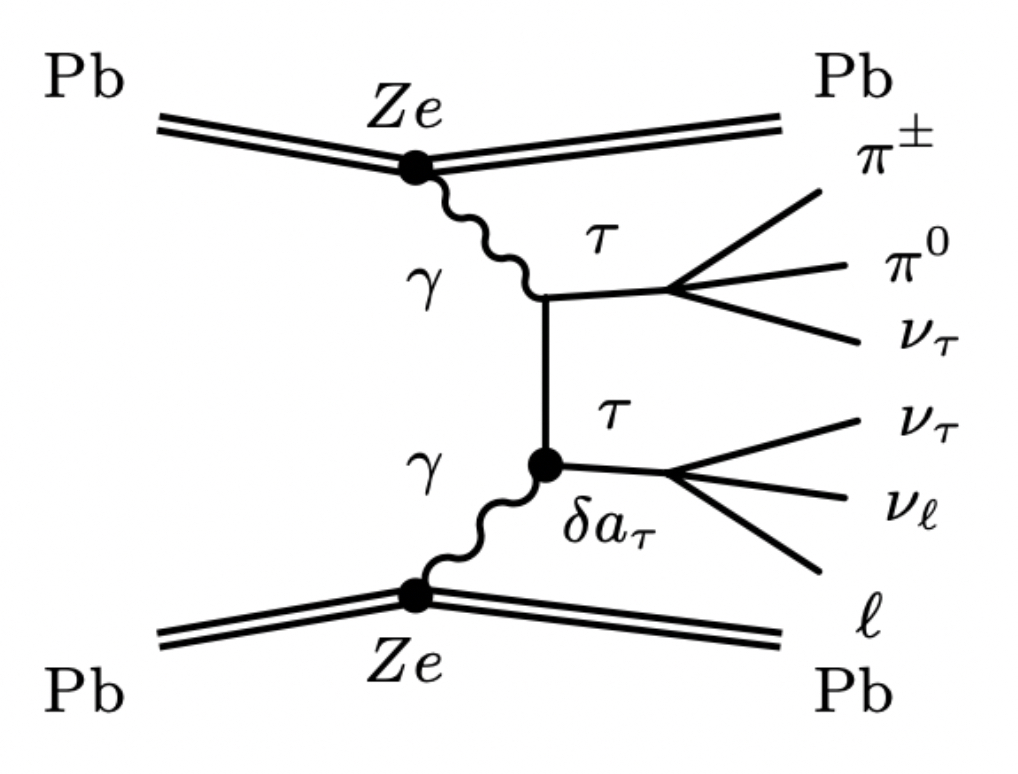}
    \caption{Pair production of tau leptons from ultraperipheral
lead ion (Pb) collisions in two  decay modes: hadronic and leptonic. New physics can affect the  tau-photon couplings modifying the magnetic moment by $\delta a_{\tau}$.}
   \label{fig:tau pairs}
    \end{figure}

 Recent papers proposed to use ultraperipheral collisions of heavy ions at LHC to measure the exclusive $\tau-pair$ production cross section \cite{Beresford} \cite{Dyndal}.
 Using Pb-Pb ultraperipheral collisions to single-out $\gamma-\gamma$ collisions yielding $\tau-pair$, offers several advantages compared with proton-proton collisions at LHC. 
 In fact, in Pb-Pb collisions the cross section for $\gamma \gamma \rightarrow \tau \tau$ (see Figure~\ref{fig:tau pairs}) is enhanced by a factor $Z^4$, largely compensating the lower integrated luminosity compared with that available in proton-proton collisions. In addition, the request of an exclusive final state containing only tau-decay products, with essentially no pile-up background, allows a better control of the background processes than in case of p-p collisions.
\par At LHC the $a_\tau$ measurements have been obtained with Pb-Pb collisions by the ATLAS and CMS experiments. CMS, with an integrated luminosity of $404.3$ $\mu b^{-1}$ obtained the  limit of $-0.088<a_{\tau}<0.056$ at 68\% CL \cite{CMSg-2}. ATLAS, using an integrated luminosity of 1.44 $\rm{nb^{-1}}$ provided a limit of $-0.057<a_{\tau}<0.024$ at 95\% CL \cite{ATLASg-2}.
These  measurements at LHC have still an uncertainty of several percents dominated by the statistical error. This uncertainty is  expected to be reduced by about one order of magnitude with the new data to be collected at HL-LHC with an increased integrated luminosity of a factor 10.

\par A measurement of  $a_\tau$ is proposed to be performed also at the new $\tau$-factories such as the  $ e^+e^- $ collider Belle2 \cite{Belle2}. It has been estimated that with 50  $ab^{-1}$ the Belle-II experiment could set the limit $|a_\tau|< 1.75 \cdot 10^{-5}$ (1.5\% of the SM prediction). 
Still at Belle-II with an integrated luminosity of 40 $fb^{-1}$ and using polarised electron beams a precision on $a_\tau$ of $10^{-6}$ could be achieved~\cite{Belle2_com}.
However no systematic uncertainty was taken into account, and the detector response was described by a fast simulation.  
\par
The theoretical prediction is: $a_{\tau}^{theo}= 117 721 (5) \times  10^{-8}$ \cite{Eidelman} where the largest contribution to the uncertainty is due to hadronic effects.  By comparing  the present $a_{\tau}^{exp}$ with $a_{\tau}^{theo}$ it is clear that the sensitivity of the  existing measurements is still more than one order of magnitude worse than needed.

\par  The  discrepancies between experiment and theory  already  observed for both $a_{e}$ and $a_{\mu}$ makes the  exploration of the tau lepton magnetic moment even more crucial for fundamental physics and more efforts should  be devoted especially in refining the experimental methods to measure it.

The growing interest around the rich physic program provided by photon interactions generated by heavy ions at LHC has also fostered the development of tools to improve the generation of these type of events. It is important to identify tools for event generation that provide a good compromise between flexibility and precision. For this reason in this work the $\tau-pair$ signal production is generated with an effective description in a UFO model (Universal FeynRules Output) \cite {UFO} implemented in the Monte Carlo generator MadGraph5 \cite{Alwall:2011uj}. This choice provides several  advantages compared with previous approaches \cite{Dyndal} allowing to distinguish the linear interference between SM and BSM and the pure quadratic BSM contribution. Moreover, an  easier and more effective interface among the particle level simulation, the showering/hadronization and the detector effects is possible. Details on the adopted model will be given in the following chapter.
 The detector performance and the experimental environment at LHC are those of the ATLAS experiment.
\par
 \par In this work the analysis of data to extract  $a_\tau$ is performed by exploiting a Gradient Boosted Decision Trees (BDTG) \cite{BDTG} approach that optimises the signal selection together with the best background rejection. To verify the performance of this new approach the results achieved with the BDTG analysis are compared with those obtained with a Standard Cuts (SC) mimicking that one applied in previous LHC experiments. Results refer to an integrated luminosity of 2.0 $nb^{-1}$ corresponding to the total integrated luminosity of the 2015-2018 heavy ions data-taking. In addition, results for 1.44 $nb^{-1}$ of integrated luminosity are also quoted to have a direct comparison with the latest published ATLAS results~\cite{ATLASg-2}.

\section{Generation of signal and background processes}

In this section the steps to generate and simulate signal and background processes are discussed. The photon flux implementation and the advantage of using Pb-Pb with respect to proton-proton collisions are discussed in section \ref{sec:flux} . The signal process $Pb(\gamma)-Pb(\gamma) \to \tau^+ \tau^-$ generation, including the contribution from BSM effects is described in section~\ref{sec:signBSM}. In this section is also discussed the effect on differential and total cross-sections due to a modified value of $a_{\tau}$. The background processes relevant to this study are presented in section~\ref{sec:bkg}. Detector effects have been simulated with a fast simulation as described in section~\ref{sec:det}.

\subsection{The photon flux}\label{sec:flux}

\begin{figure}[htbp]
    \centering
    \includegraphics[width=0.48\linewidth]{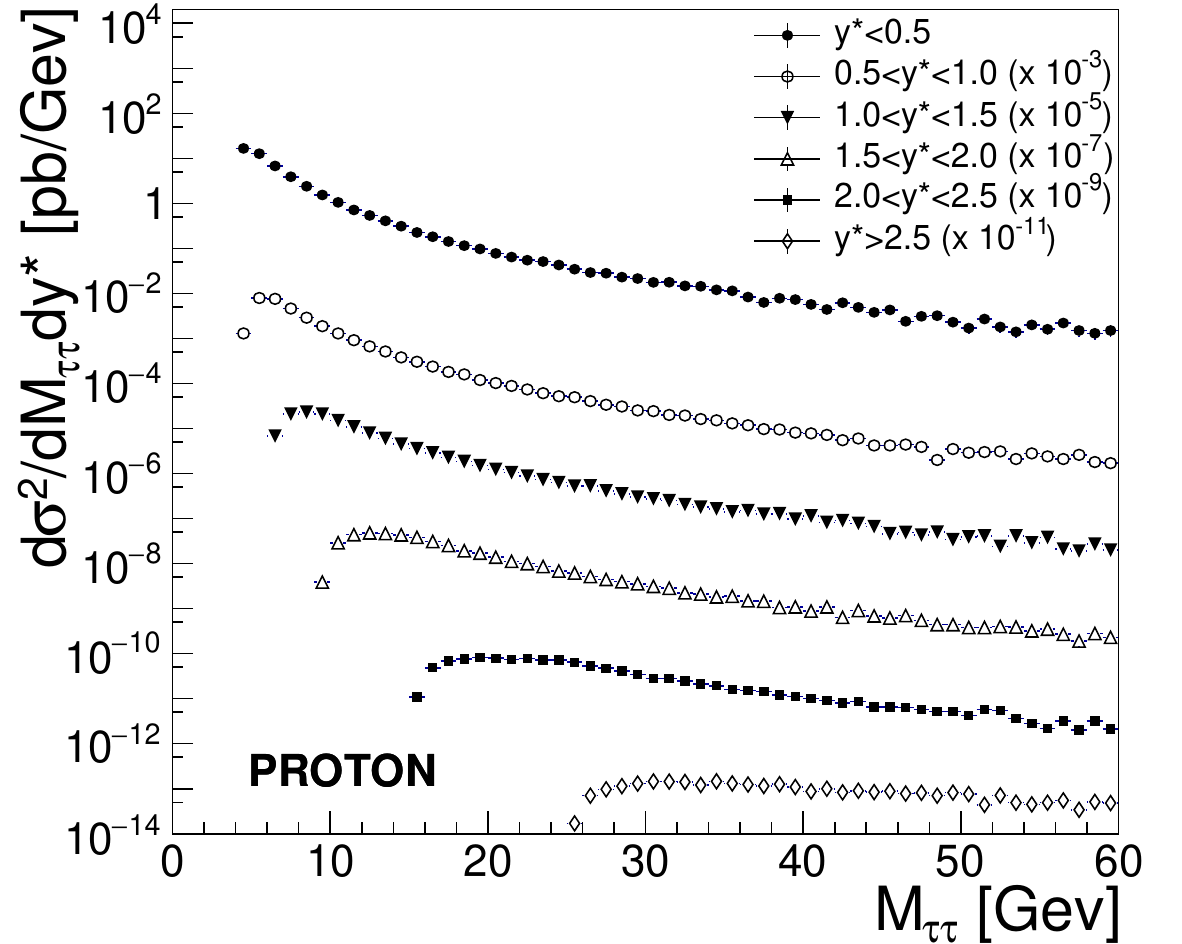}
    \includegraphics[width=0.48\linewidth]{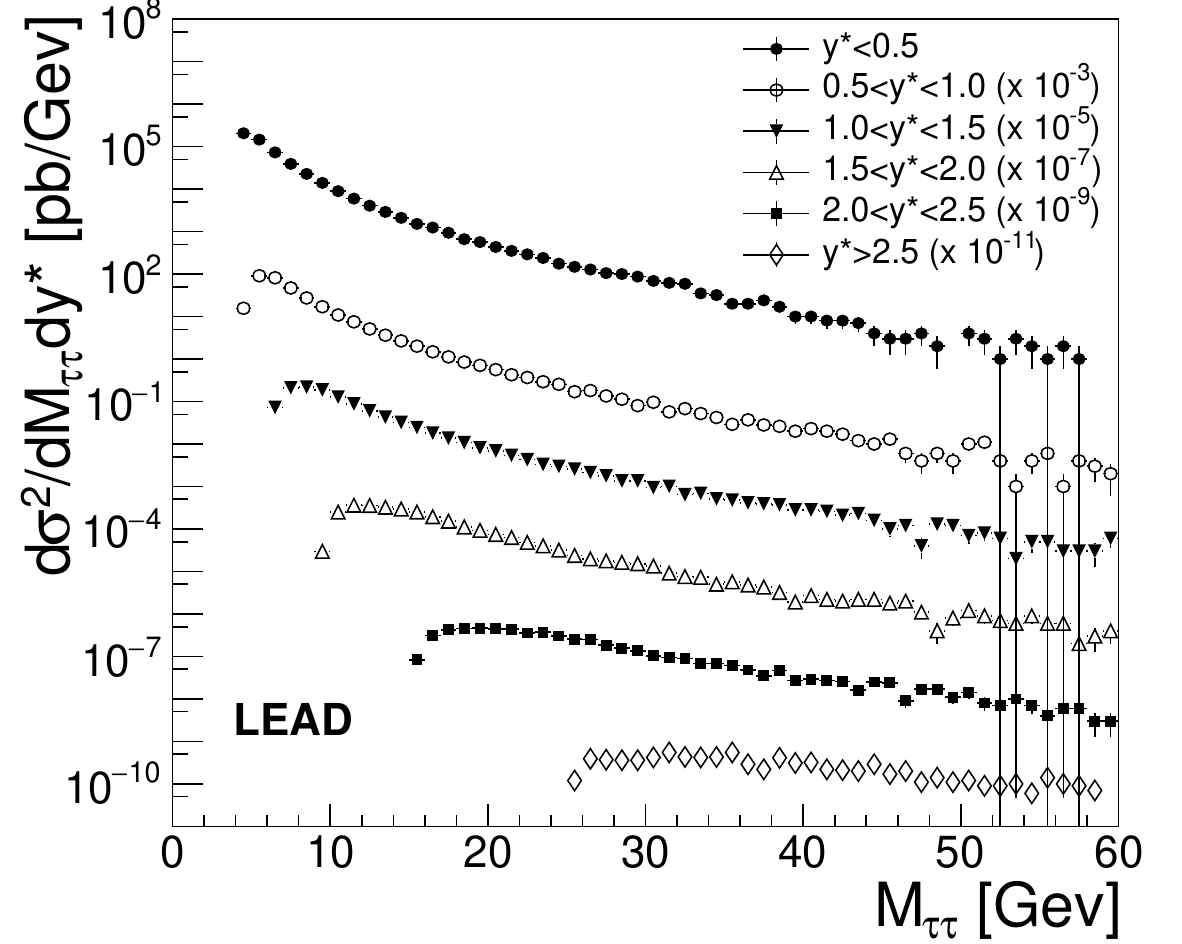}
   
    \caption{ 
   Double-differential di-tau cross sections as a function of the di-tau mass in bins of half the rapidity separation, $y^{*}$ ($y^{*} = |y_1-y_2|/2$), of the 2 taus. The cross sections for proton (left plot) and lead (right plot) photon flux are shown.  
 }
    \label{fig:photflux}
\end{figure}

\begin{figure}[htbp]
    \centering
    \includegraphics[width=0.48\linewidth]{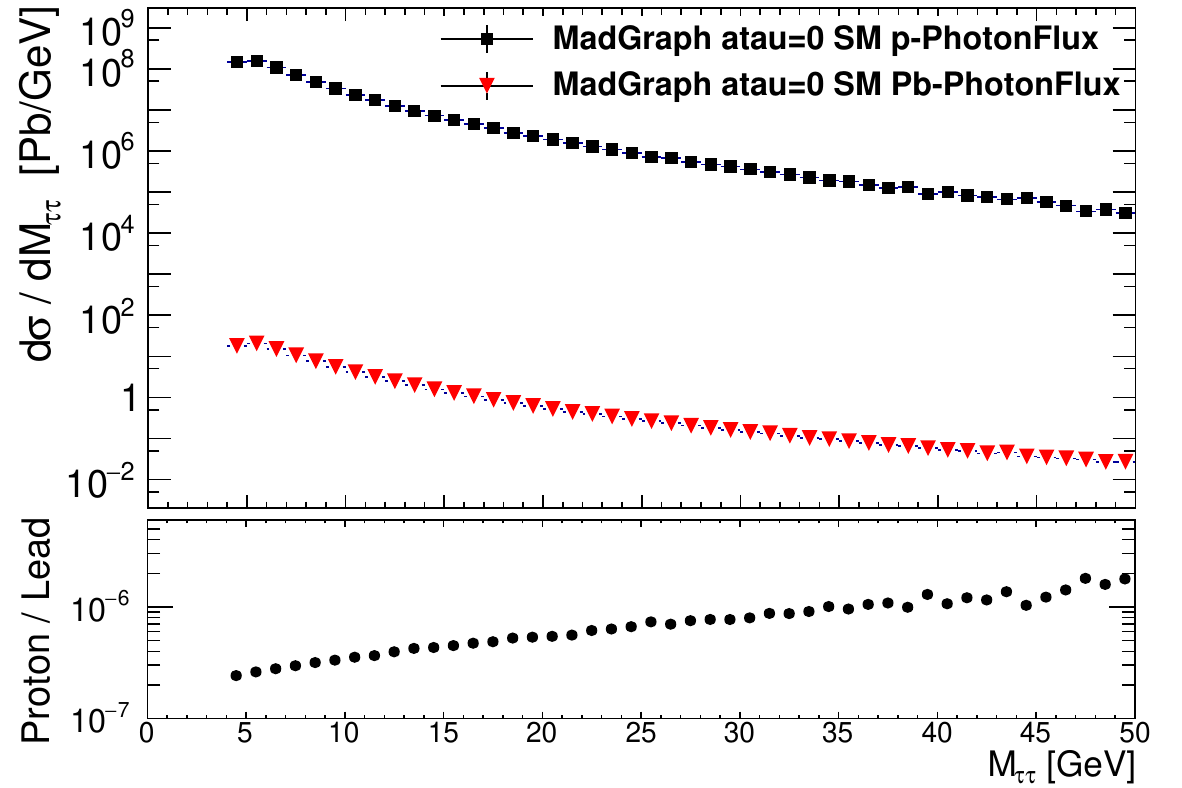}
    \includegraphics[width=0.48\linewidth]{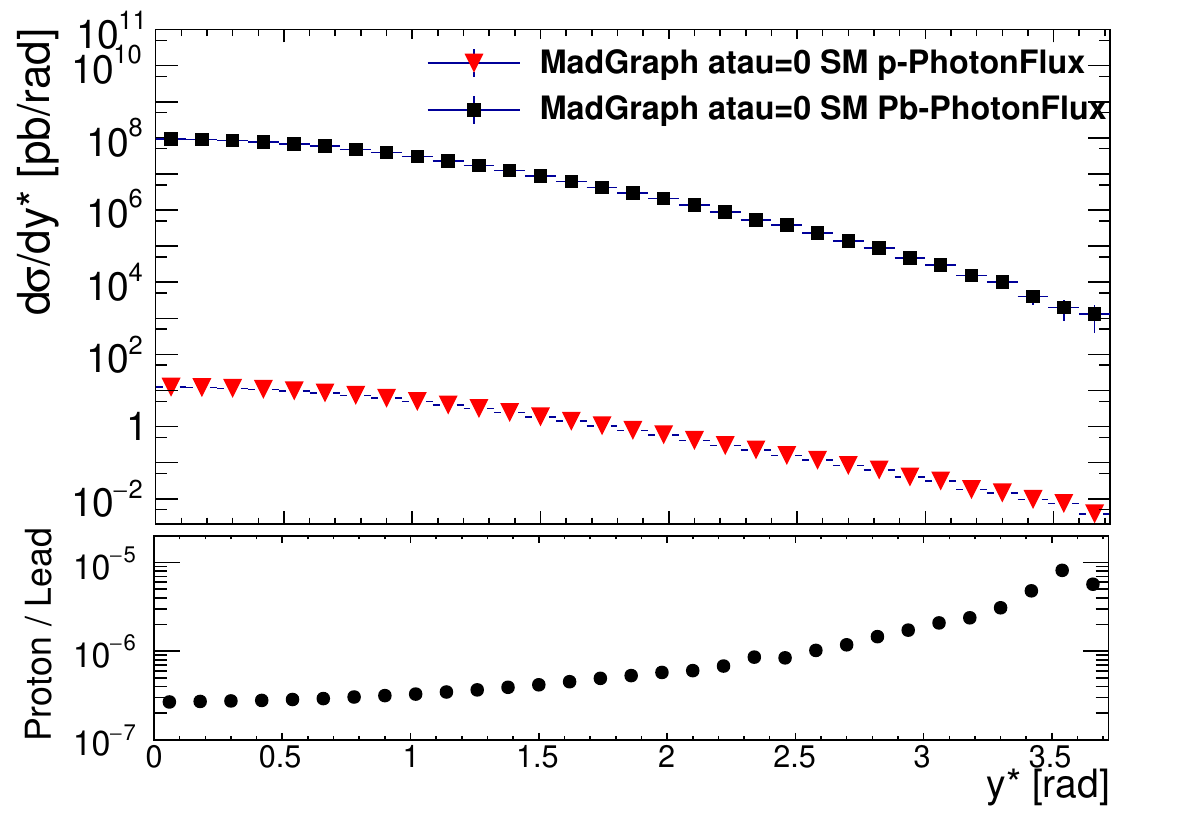}
   
    \caption{ 
   Differential di-tau cross sections as a function of the di-tau mass (left plot) and as a function of rapidity separation, $y^{*}$ ($y^{*} = |y_1-y_2|/2$), of the 2 taus (right plot) depending on the photon flux. The ratio of the cross section proton over lead is reported at the bottom of each plot.
 }
    \label{fig:photppbflux}
\end{figure}

In this work the process $Pb(\gamma)-Pb(\gamma) \to \tau^+ \tau^-$ is generated by modifying MadGraph5 to include the photon flux from the lead beams in ultra-peripheral collisions following the prescription in reference~\cite{dEnterria:2009cwl}. In the Equivalent Photon Approximation (EPA) \cite{vonWeizsacker:1934nji,Williams:1934ad}, and neglecting non-factorizable hadronic interactions between nuclei and nuclear overlap effects, the $\gamma \gamma \to \tau^+ \tau^-$ cross section in ultra-peripheral Pb-Pb collisions can be expressed as the convolution:
\begin{equation}\label{eq:xsec}
\sigma^{(Pb-Pb)}(\gamma \gamma \to \tau^+ \tau^-) = \int dx_1 dx_2 N(x_1) N(x_2) \hat{\sigma}(\gamma \gamma \to \tau^+ \tau^-) \, ,
\end{equation}\label{eq:flux}
where $\hat{\sigma}(\gamma \gamma \to \tau^+ \tau^-)$ is the elementary cross-section and $N(x_i)$ represents the photon flux from the two Pb-ions, calculated as a function of the ratio of the emitted photon energy from the ion $i$ with the beam energy ($x_i= E_i/E_{\text{beam}}$). $N(x_i)$ is described by the classical analytic form \cite{Jackson:1998nia}:
  \begin{align}\label{eq:photon-flux}
  N(x_i)= & \frac{2 Z^2 \alpha}{x_i \pi} \big\{ \bar x_i K_0(\bar{x_i} )K_1(\bar{x_i} ) - \frac{\bar{x_i}^2}{2} [K^2_1(\bar{x_i}) - K^2_0(\bar{x_i}) ] \big\} \\ \nonumber
  & x_i= E_i/E_{\text{beam}}\, , \quad \bar x_i= x_i\, m_N b_{\text min}/2  \, ,
  \end{align}
  where, for Pb, $Z=82$, $A=208$, the nucleon mass $m_N= 0.9315$~GeV, the nucleus radius $R_A\simeq 6.09 A^{1/3}$~GeV$^{-1}\simeq 7$~fm, $b_{\text min} \simeq 2 R_A$ is the minimum impact parameter and $K_0 (K_1)$ are  the modified Bessel functions of the second kind of the first (second) order. The same implementation of the photon flux is also used in \cite{Beresford}, where it is found that a more complete treatment of nuclear effects, as included in program as SUPERCHIC \cite{Harland-Lang:2018iur}, do not impact significantly the cross sections and distributions of the processes which are relevant for our study.  \\
A comparison between the di-tau double differential cross section from proton-proton and from lead-lead collisions is shown in Figure \ref{fig:photflux}.  The proton distributions are obtained using MadGraph5 default configuration, which adopts the EPA improved Weizsaecker-Williams formula \cite{Budnev:1975poe}. The figures show the double-differential di-tau cross sections as a function of the di-tau mass in bins of half the rapidity separation, $y^{*}$ ($y^{*} = \frac{|y_1-y_2|}{2}$), of the 2 taus. 
\\The comparison between lead and proton cross sections and their ratio as a function of di-tau mass and of the di-tau rapidity integrated on the di-tau separation and di-tau mass respectively, are shown in Figure \ref{fig:photppbflux}. It is interesting to note that the expected $Z^4$ enhancement in favour of the  radiation intensity from Pb reduces as di-tau mass or rapidity separation  increases. In fact, as the di-tau mass (or the di-tau separation) increases, also the $Q^2$ of the interaction increases, and the interaction radius decreases accordingly. In this situation, the electromagnetic form factor generated by the lead nucleus decreases its  effectiveness in  photon emission.   This effect is encoded in the photon flux dependence on $\bar x$ of the analytic form in Eq.~(\ref{eq:photon-flux}), which is based on classical electrodynamics. 
\par
\begin{figure}[htbp]
    \centering
    \includegraphics[width=0.6\linewidth]{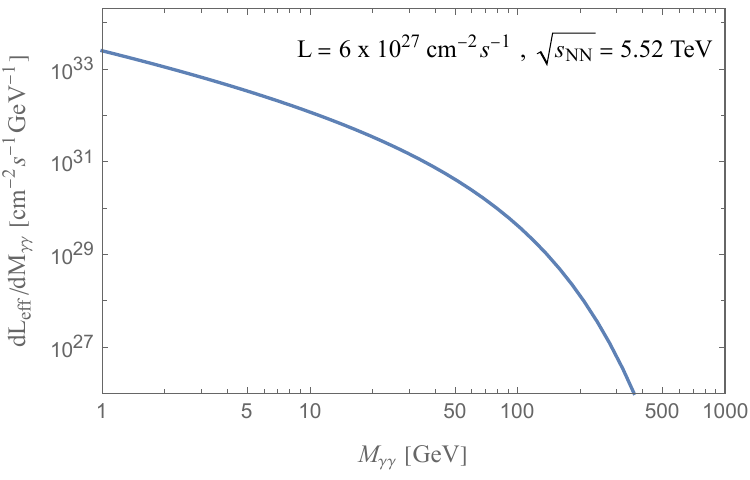}
   
    \caption{ 
  Effective $\gamma\gamma$ luminosity vs. photon-fusion mass in ultra-peripheral Pb-Pb collision at $\sqrt{s_{NN}}=5.52$ TeV.
 }
    \label{fig:photoflux}
\end{figure}

 The cross section in Eq. (\ref{eq:xsec}) can be also expressed in terms of an effective $\gamma\gamma$ luminosity ($\frac{d L_{eff}}{d M_{\gamma\gamma}}$) as:

\begin{equation}\label{eq:xsec-W}
\sigma^{(Pb-Pb)}(\gamma \gamma \to \tau^+ \tau^-) = \int dM_{\gamma\gamma} \frac{d L_{eff}}{d M_{\gamma\gamma}} \hat{\sigma}(\gamma \gamma \to \tau^+ \tau^-) \, .
\end{equation}
Figure \ref{fig:photoflux} shows the effective $\gamma\gamma$ luminosity, as a function of the photon-fusion mass $M_{\gamma\gamma}$, as obtained from the convolution of the photon flux in (\ref{eq:flux}).

\subsection{Generation and simulation of the signal}\label{sec:signBSM}
Events including BSM physics through a modified value of $a_\tau$ are generated implementing in a UFO model \cite{Degrande:2011ua}, to be used in MadGraph5, the effective Lagrangian term:

\begin{equation}
\mathcal{L}_{a_\tau}= a_\tau \frac{e}{4 m_\tau} \bar \tau_L \sigma^{\mu\nu}\tau_R F_{\mu\nu} \, ,
\end{equation}
by means of  Feynrules \cite{Alloul:2013bka}. The implementation is validated against theoretical analytical predictions and previous results from LEP \cite{DELPHI:2003nah}.  \\

The approach to generate BSM effects here described differs from previous analysis. In fact, the authors in \cite{Dyndal} use a custom code, which generates the signal by means of the full form of the photon-tau vertex function and of the cross sections calculated at leading order. The MadGraph5 approach, implementing the signal generation via an effective description in a UFO model, allows an easier interface with showering/hadronization effects and with the detector simulation. Moreover, it also allows to easily single-out the linear interference terms with the SM from the purely BSM quadratic terms. The SM and BSM $\gamma\gamma \to \tau^+ \tau^-$ inclusive cross sections here obtained show an agreement within 10\%  with those in \cite{Dyndal}.

\par
The study in \cite{Beresford} adopts our same implementation of the photon flux in MadGraph5, as also  MadGraph5 for signal simulations. However, \cite{Beresford} makes use of a different UFO model, the SMEFTsim package \cite{Brivio:2017btx}, and extracts the BSM modification to $a_{\tau}$ from the parameters of the SM effective field theory considered in this SMEFTsim code. 

\par A significant discrepancy is observed between the BSM signal cross section values of \cite{Beresford} and the calculation here presented; on the contrary the two SM results are in agreement. Since the two BSM calculations rely on an EFT approach, the source of the disagreement it is most probably not connected to the EFT but to an issue occurred in \cite{Beresford} with the conversion between SMEFTsim operators and those generating modification to $a_\tau$. A similar discrepancy is observed between the results in \cite{Beresford} and the BSM cross section calculations reported in \cite{Dyndal}.

\begin{figure}[htbp]
\centering

\includegraphics[width=0.7\linewidth]{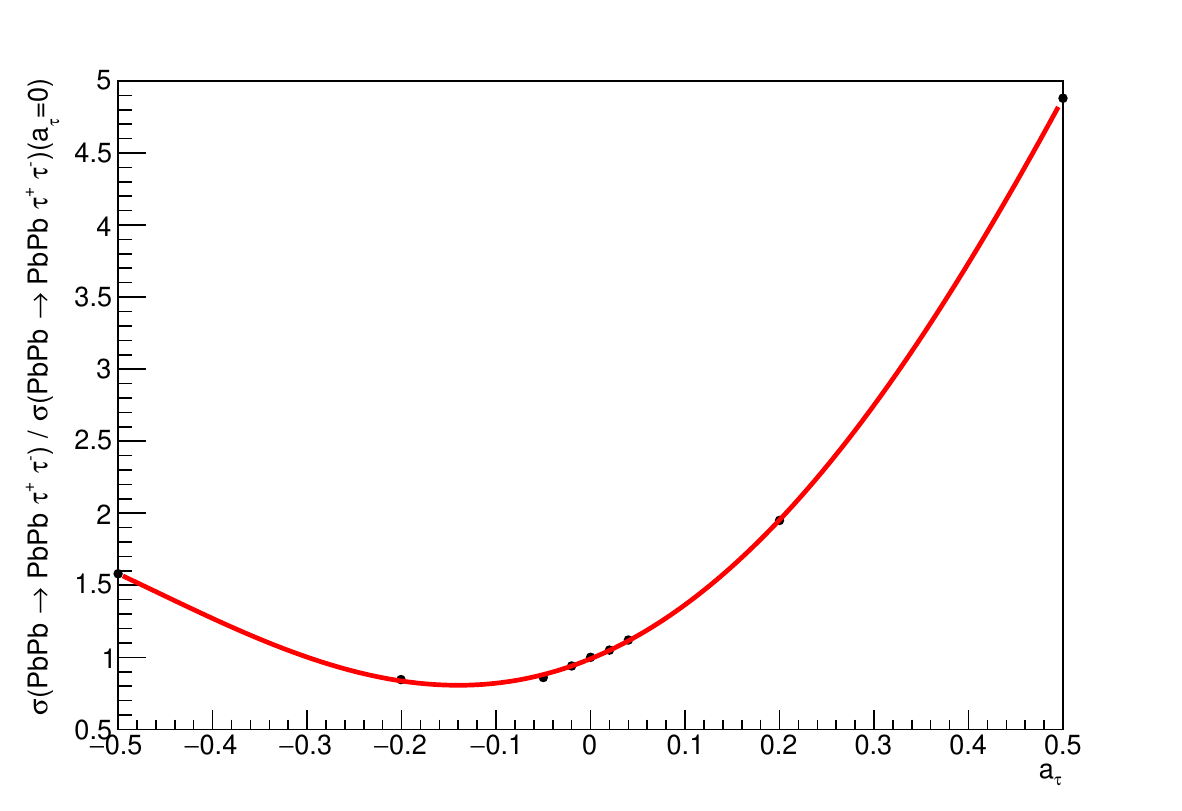}
  \caption{Ratio between the total ultra peripheral  cross sections for $Pb(\gamma)-Pb(\gamma) \to \tau^+ \tau^-$ production at the LHC energy $\sqrt{s_{NN}}=5.02$ TeV and the SM cross section (\atau= 0) as a function of \atau. At generation level a cut on lepton ${ p_{T} >1 \, GeV}$ is applied.
}
\label{fig:atau_vs_xsec}
\end{figure}

\begin{figure}[htbp]
    \centering
    \includegraphics[width=0.48\linewidth]{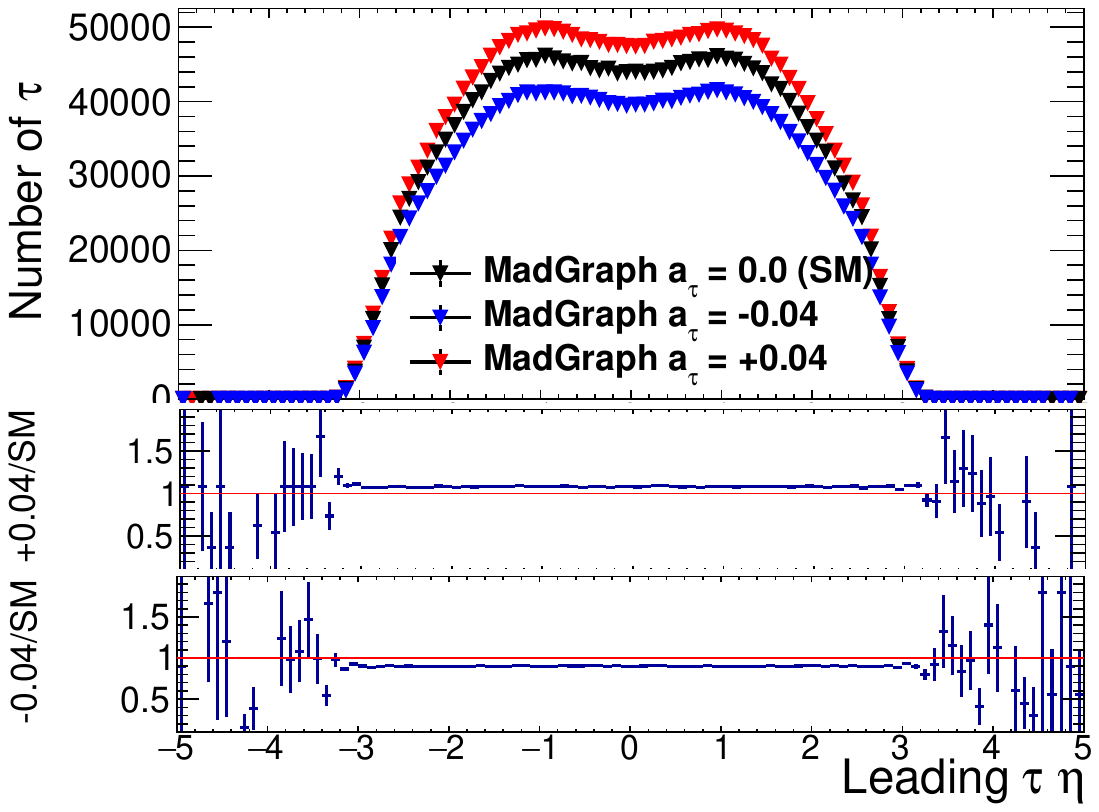}
    \includegraphics[width=0.48\linewidth]{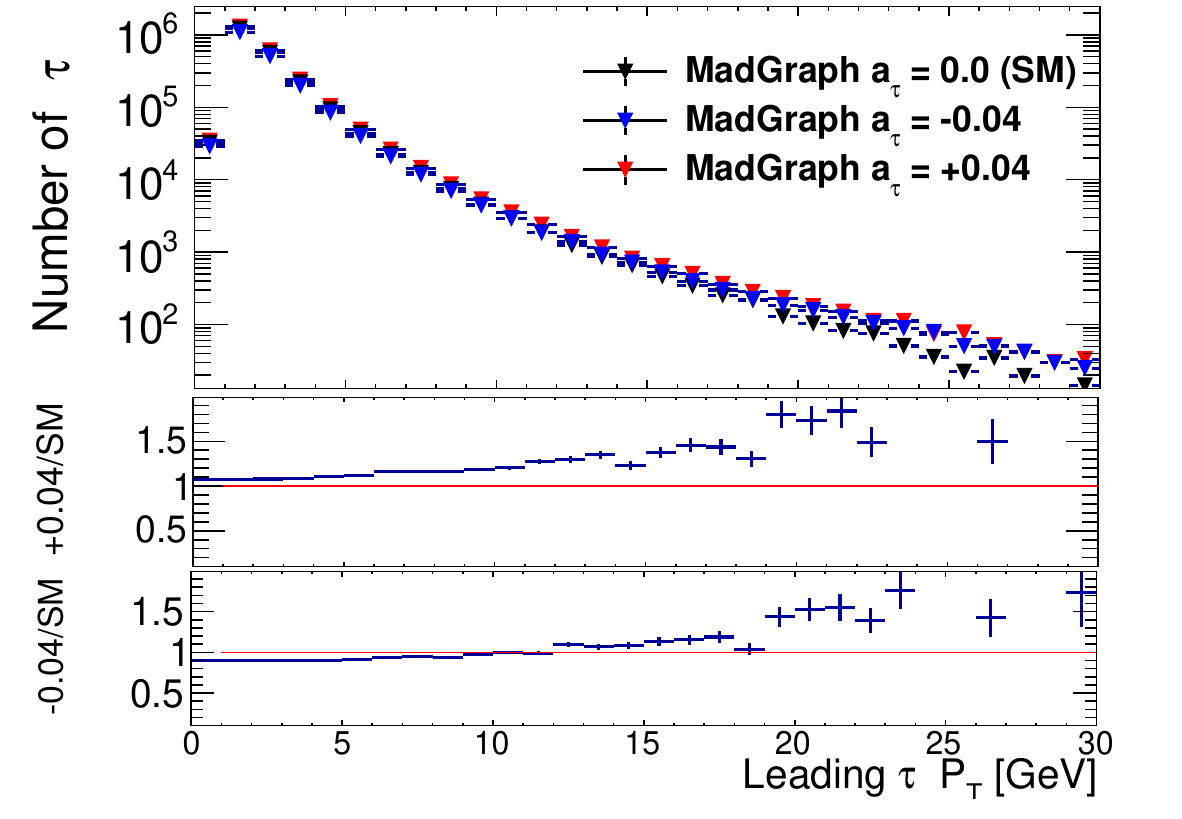}
   \includegraphics[width=0.48\linewidth]{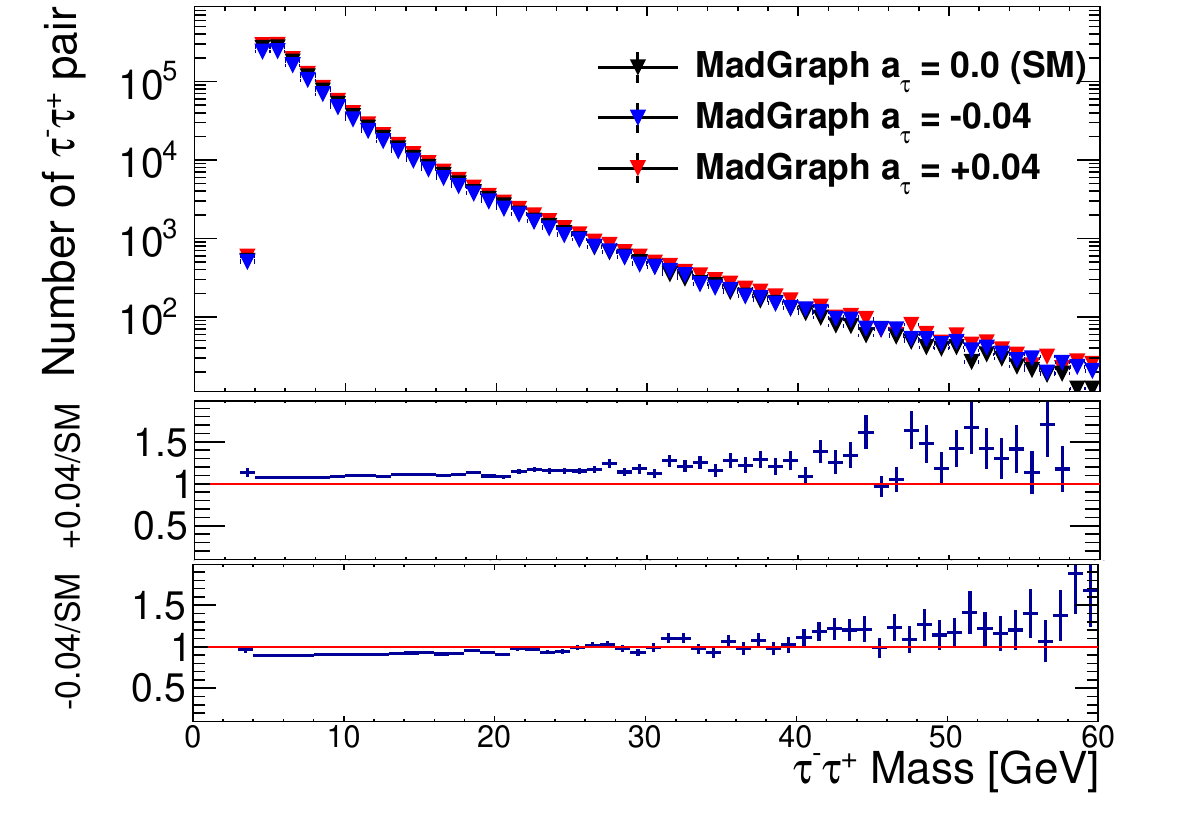}
    \includegraphics[width=0.48\linewidth]{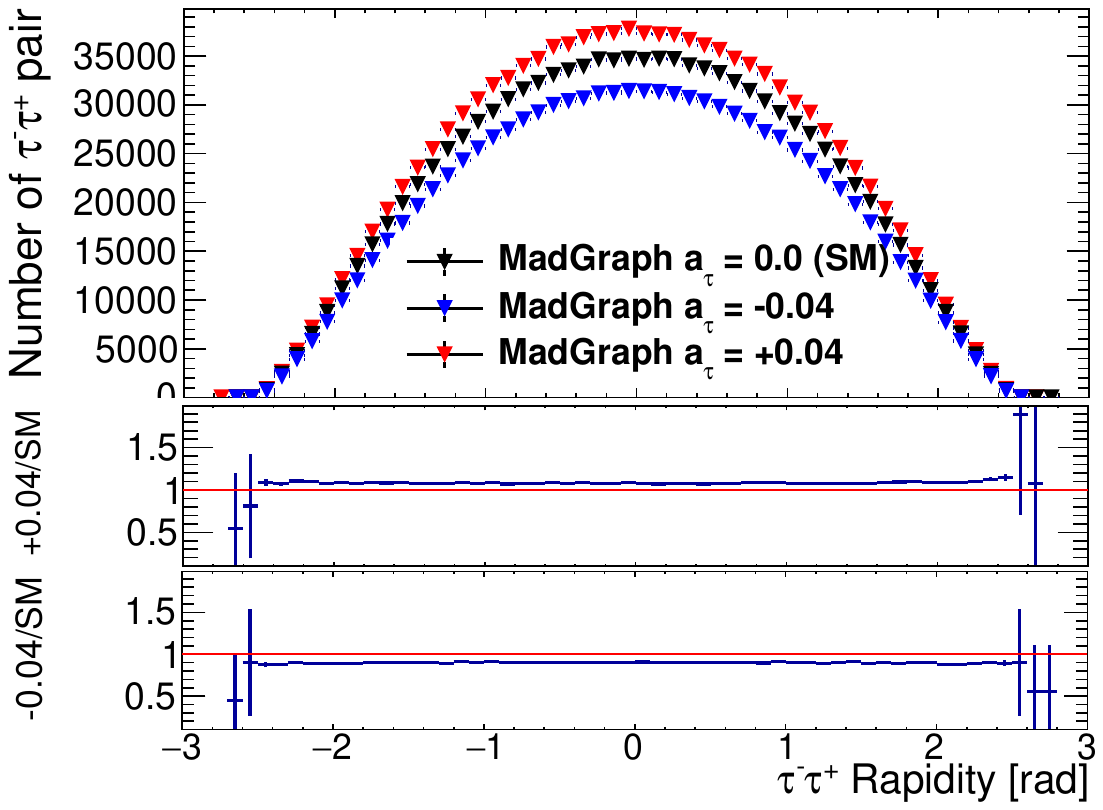}
    \caption{Top plots: Leading-tau $\eta$ and $p_T$ distributions for different values of $a_{\tau}$: +0.04, -0.04, 0. The ratio between $a_{\tau}$ =$\pm$0.04 and $a_{\tau}$ =0 is reported in the bottom side of each plot. Bottom plots:  di-tau system mass and rapidity distributions at different values of $a_{\tau}$: +0.04, -0.04, 0. The ratios are reported in the bottom side of the plots.}
    \label{fig:at_004}
\end{figure}

The ratio between the $Pb(\gamma)-Pb(\gamma) \to \tau^+ \tau^-$ total cross section and the SM cross section as a function of $a_\tau$ is shown in Figure \ref{fig:atau_vs_xsec}, where the ratio is set to 1 for $a_\tau = 0$, considered as the SM value. The asymmetry between positive and negative \atau values is due to interference between the SM part and the BSM modified $\tau$ coupling.
The effect of different \atau values is investigated by looking at various $\tau$ and di-$\tau$ kinematical distributions. In particular,
Figure~\ref{fig:at_004} shows the leading $\tau$ $p_{T}$, the leading $\tau$ rapidity, the di-tau system rapidity and invariant mass distributions for three representative values of \atau ($0,\pm 0.4$) normalized to $2 \rm{nb^{-1}}$ of integrated luminosity. Figure~\ref{fig:at_004} proves that, in addition to the $\tau$-pair cross section, also the differential cross sections, especially the $\tau$ $p_T$ distribution, can be exploited to improve the sensitivity to $a_\tau$.

The $\tau$ decays, the hadronization and the shower processes are described with PYTHIA8 \cite{pythia}.
Two millions events have been generated for each signal sample, varying the coupling $a_{\tau}$ from -0.04 to +0.04 (see appendix~\ref{MonteCarlo} for the complete list and statistics). 
\subsection{Background processes}~\label{sec:bkg}
The requirement of selecting exclusive di-tau decay products in UPC events greatly reduces the background contribution in the signal selection. The background processes considered are $\gamma \gamma \rightarrow e^{+} e^{-}$, $\gamma \gamma \rightarrow \mu^{+} \mu^{-}$,$\gamma \gamma \rightarrow  b \bar{b}$, 
$\gamma \gamma \rightarrow  jet(c,s,u,d) jet(\bar{c},\bar{s},\bar{u},\bar{d})$. 
Among these processes the $\gamma \gamma \rightarrow \mu^{+} \mu^{-}$ processes where one of the $\mu$ radiate a photon is the major source of background. 
As shown  in \cite{Beresford} and \cite{Dyndal}, the $\gamma \gamma \rightarrow\bar{q}q$
produces a larger charged-particle multiplicity than the signal and hence can be totally rejected by exclusivity requirements.

Other contributions to the background could be due to diffractive photo-nuclear events, mediated by a Pomeron exchange, where the Pb-ions may not dissociate and some particles could be  produced in the central rapidity region. For  this  background,  a reliable Monte Carlo simulation is not available, however in reference \cite{ATLASg-2} it was estimated, by a data-driven method, that this contribution results in a about $2\%$ contribution to the $\tau^+ \tau^-$ data sample. In this analysis this contribution has been neglected.

Two millions events of background samples have been produced with PYTHIA8, see appendix~\ref{MonteCarlo} for details.

\subsection{Simulation of detector effects}~\label{sec:det}
The simulation of the ATLAS detector is done by using DELPHES 3.5.0 framework \cite{DELPHES}. This package implements a fast-simulation of the detector, including a track propagation system embedded in a magnetic field, the electromagnetic and hadron calorimeter responses, and a muon identification system. Physics objects as electrons or muons are then reconstructed from the simulated detector response using dedicated sub-detector resolutions. For the analysis here presented, electron \cite{Eff_ele} and muon \cite{Eff_muon} efficiencies have been modified using the latest ATLAS performance results as obtained on the data sample collected in 2015-2018 (Table~\ref{tab:eff_delphes}).
Other efficiencies such as the tracking efficiency or the smearing reconstruction functions are used without changes.                       		    
\begin{table}[ht!]
\begin{center}
    \begin{tabular}{|c|c|c|}
    \hline
      Particle& $\eta$ and $p_{T}$ [GeV]  & Efficiency [$\epsilon$] \\
       \hline
        \multirow{5}{*}{Electron } &  &\\     
    &$|\eta|>2.4$ and $p_{T}<=4.5$ & 0.00\\
    &$|\eta|<=2.4$ and $4.5 >p_{T}<30.0$  & 0.82\\
    &$|\eta|<=2.4$ and $30.0 >p_{T}<40.0$   & 0.86\\
    &$|\eta|<=2.4$ and $40.0 >p_{T}<=60.0$   & 0.88\\
    &$|\eta|<=2.4$ and $p_{T}>60.0$   & 0.92\\
    \hline
    \multirow{4}{*}{Muon } &  &\\   
    &$p_{T}<=3.5$ ~GeV & 0.00\\
    &$|\eta|<=2.5$ and $3.5 >p_{T}<4.0$   & 0.65\\
    &$|\eta|<=2.5$ and $4.0 >p_{T}<5.0$   & 0.80\\
    &$|\eta|<=2.5$ and $p_{T}>5.0$   & 0.95\\
  
        \hline
        
        \hline
    \end{tabular}
\end{center}
\caption{\small Tracking efficiencies, as applied in DELPHES, for electrons \cite{Eff_ele} and muons \cite{Eff_muon} for different  $\eta\times p_{T}$ bins.} \label{tab:eff_delphes}
\end{table}

\section{Analysis procedure}
In this section the procedure to select the signal from the background processes is described. The analysis is applied to the data including the fast detector simulation.  The preselection cuts and the signal region definition are described in section~\ref{sec:pre}. The two analysis procedures based on Standard Cuts (SC) and on a multivariate approach (BDTG) respectively are presented in section~\ref{sec:sig}. 

\subsection{Preselection and signal region definition}\label{sec:pre}
Event selection requires at least one $\tau$ decayed leptonically. The second $\tau$ is requested to decay hadronically and is reconstructed requiring one or three tracks. Two signal regions are identified according to the $\tau$ decay topologies: one lepton one track (1L1T) and one lepton three tracks (1L3T) respectively. These requirements potentially collect about $22\%$  of all possible $\tau$-pair decays, as shown in Table~\ref{tab:taudecays}. The signal region where both the $\tau$'s decay leptonically is not included in this analysis due to the low statistics obtained after the lepton identification, see Table~\ref{table:2lep} in appendix~\ref{cuts} for more details.  
The final state with leptons is fundamental for the trigger selection. 
\begin{table}[ht!]
\begin{center}
\begin{tabular}{|l|c|c|}
  \hline
     Tau Decay & $\tau$ Decay Process&  Branching \\
      Definition & &   Fraction \\
    \hline
   \small{Lepton Decay} &$\tau^{-}\rightarrow e^{-}\Bar{\nu}_{e}\nu_{\tau}$ & 17.85\%\\
    &$\tau^{-}\rightarrow \mu^{-}\Bar{\nu}_{\mu}\nu_{\tau}$ & 17.36\%\\
    
    \small{One Charged Pion Decay} &$\tau^{-}\rightarrow\pi^{-}\nu_{\tau}n\pi^{0}$ & 46.75\% \\
    \small{} &(n=0,1,2,3)&\\
    \small{Three Charged Pion Decay} &$\tau^{-}\rightarrow2\pi^{-}\pi^{+}\nu_{\tau}n\pi^{0}$ & 13.91\%  \\
    \small{} &(n=0,1)&\\
    \hline
\end{tabular}
\caption{$\tau$ decay branching fractions.\label{tab:taudecays}}
\end{center}
\end{table}

\begin{figure}[htbp]
    
    \includegraphics[width=0.49\linewidth]{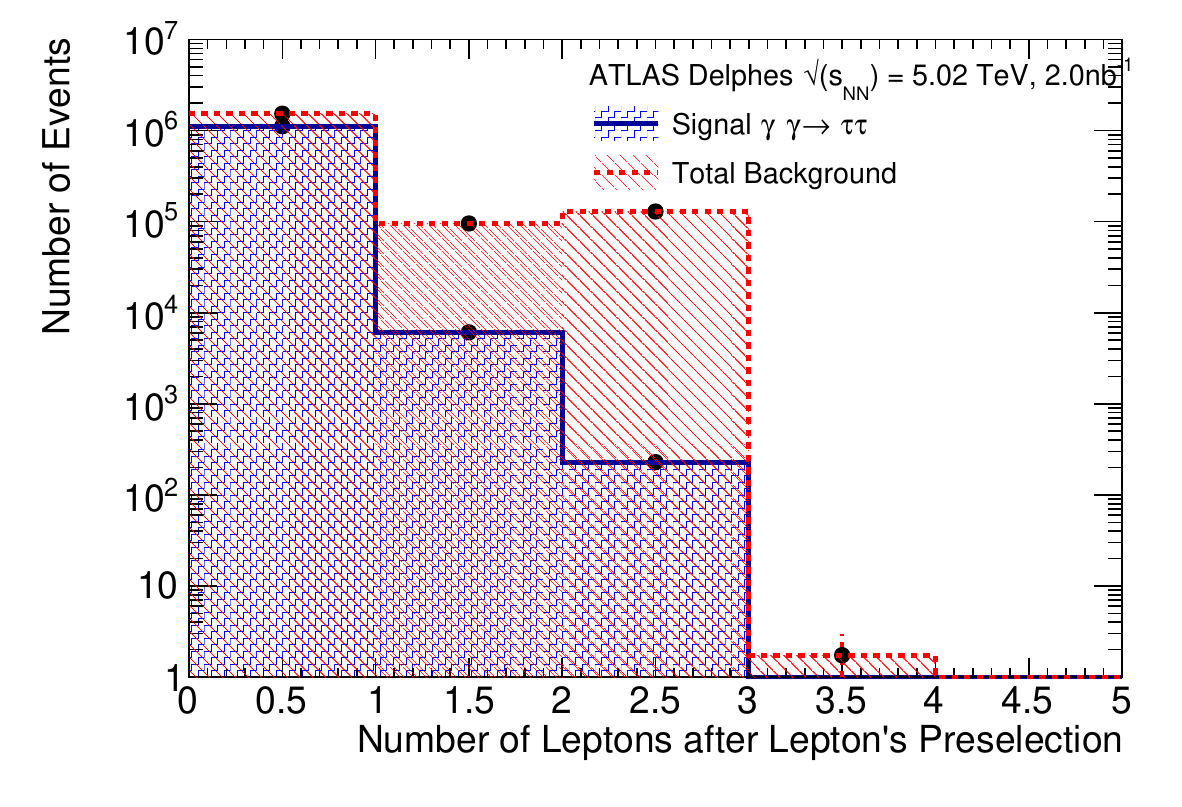}
    \includegraphics[width=0.49\linewidth]{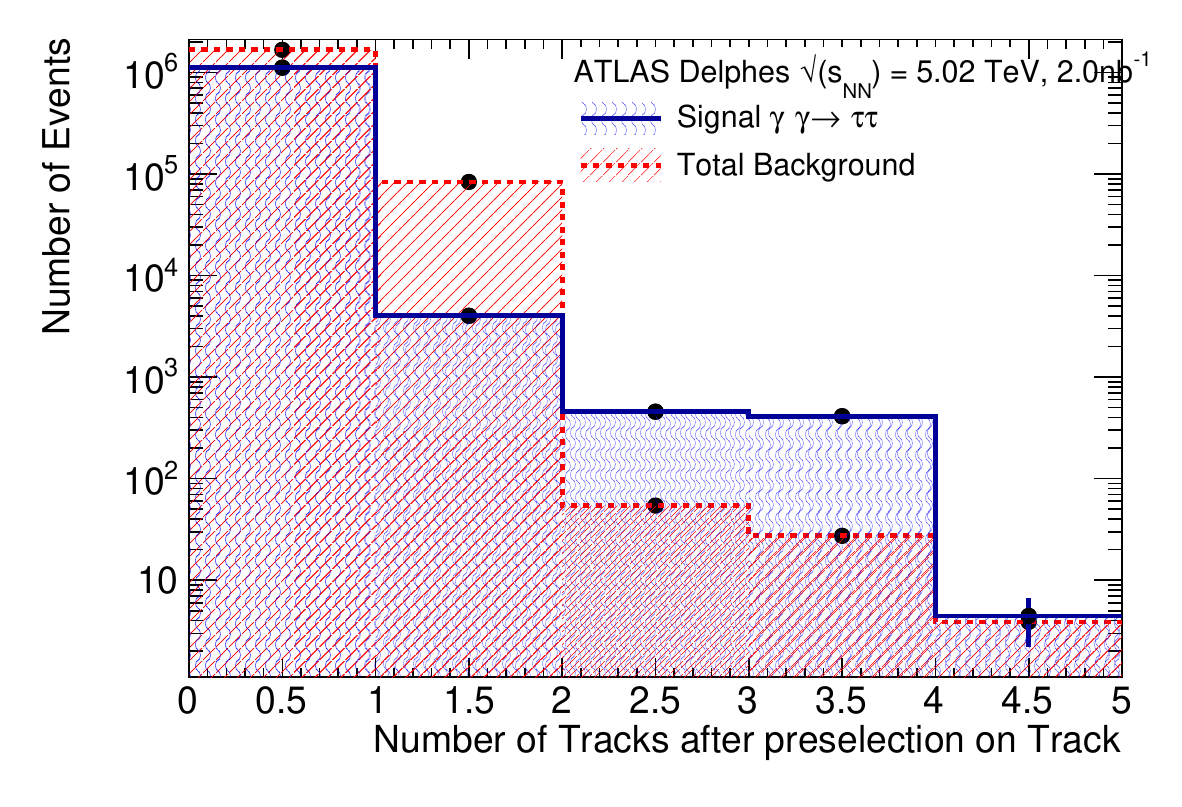}
 \caption{ 
    \label{fig:number} Lepton (left) and track (right) multiplicities for signal and background processes after the preselection cuts.}
\end{figure}

Preselection cuts, mimicking the minimal ATLAS object selection, are applied on leptons: ${p_T >4.5~(3.5)}$ GeV and $\rm{|\eta|<2.4~(2.5)}$ for electrons (muons).
In addition, each track is requested to satisfy minimal acceptance criteria: ${p_T^{(track)}>}$ 500 MeV and $\rm{|\eta^{(track)}|<}$ 2.5. The preselection cuts are summarized in Table~\ref{tab:preselection}.
The lepton and track multiplicities for signal and background processes after the preselection cuts are shown in Figure~\ref{fig:number}, the plots show that a requirement of a single lepton and one or three tracks collect most of the signal events rejecting a large fraction of the background. 
\begin{table}[ht!]
\begin{center}
\begin{tabular}{|l|c|}
\hline
 \hline
Preselection & Cuts\\
     \hline
     Electron Identification & ${p_T >4.5}$ GeV,~ $\rm{|\eta|<2.4}$\\
     Muon Identification & ${p_T >3.5}$ GeV,~ $\rm{|\eta|<2.5}$ \\
     Track Identification  & ${p_T^{(track)}>}$ 500 MeV, ~$\rm{|\eta^{(track)}|<}$ 2.5\\
    \hline
  \hline
\end{tabular}
\caption{Preselection cut summary.}
\label{tab:preselection}
\end{center}
\end{table}

\subsection{Signal extraction: SC and BDTG selections}\label{sec:sig}   
\begin{figure}[htbp]
    
    \includegraphics[width=0.49\linewidth]{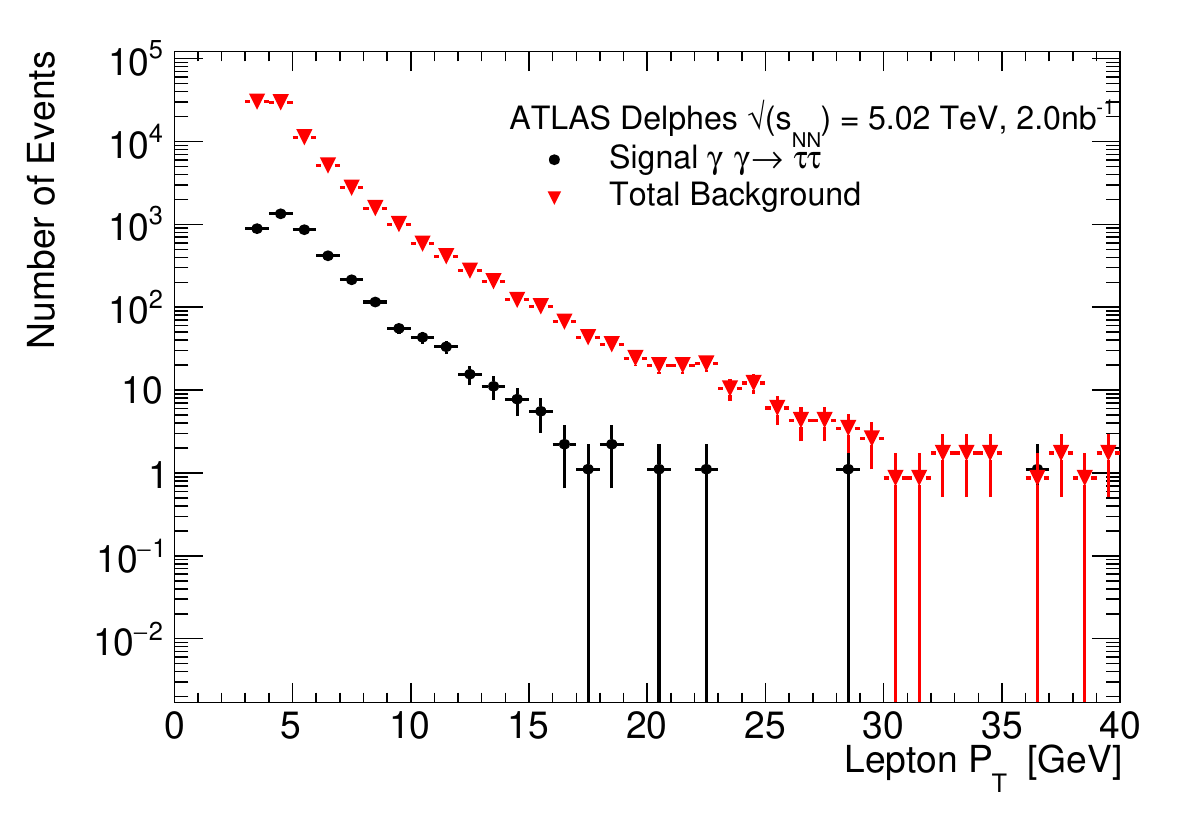}
    \includegraphics[width=0.49\linewidth]{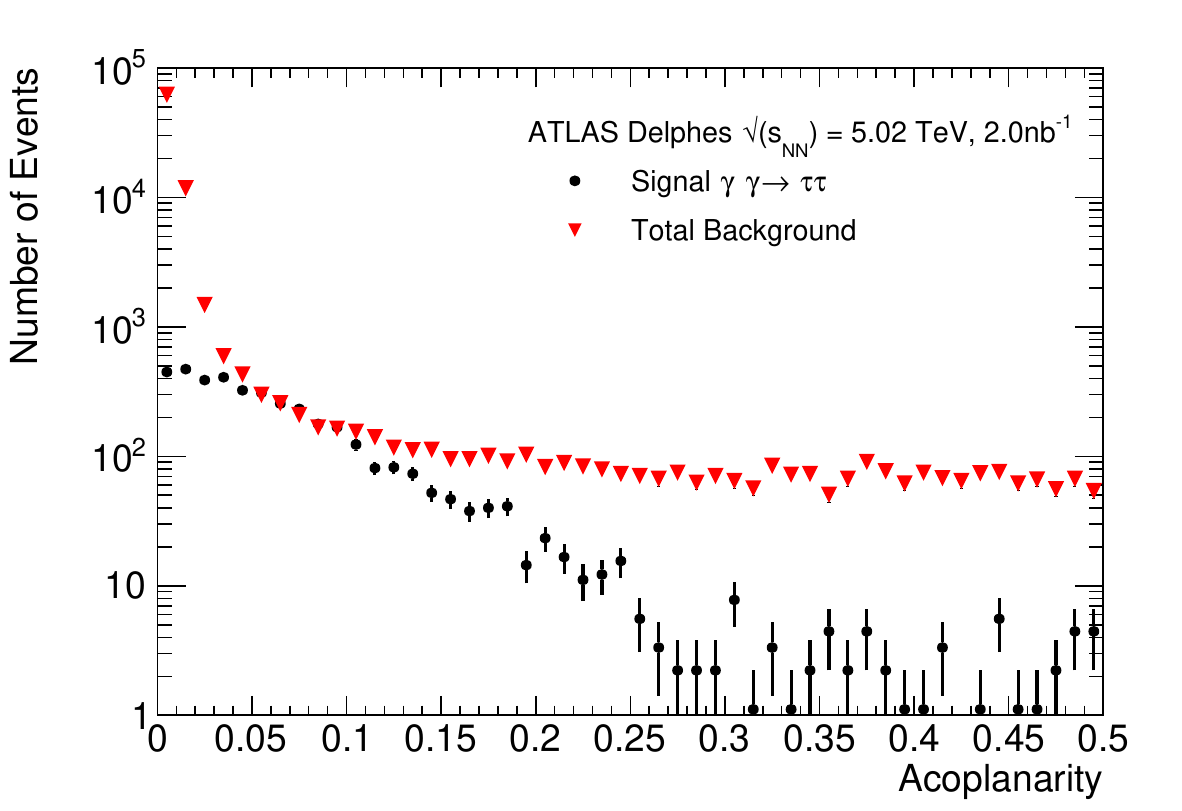}
 
 \caption{ 
    \label{fig:distribution} Distributions of the leading lepton $p_{T}$ and of the acoplanarity after the preselection for the signal region 1L1T.}
\end{figure}

\begin{figure}[htbp]
    
    \includegraphics[width=0.49\linewidth]{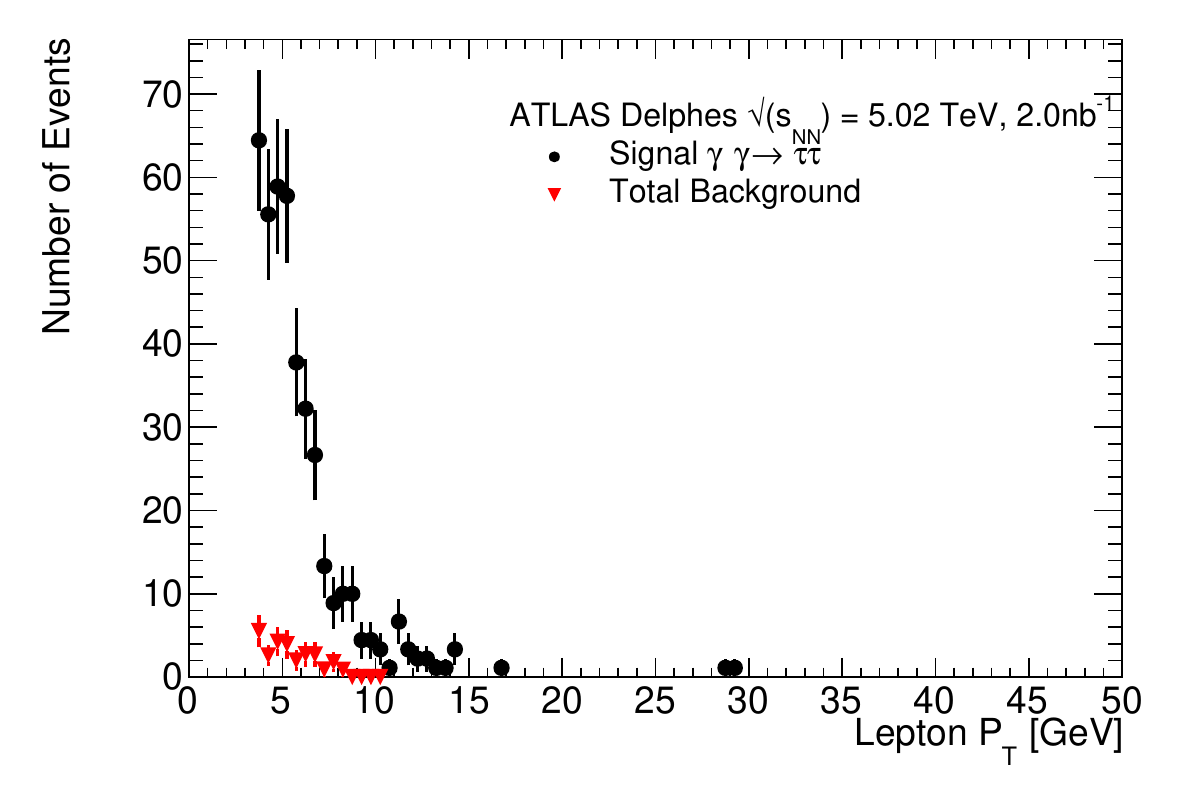}
    \includegraphics[width=0.49\linewidth]{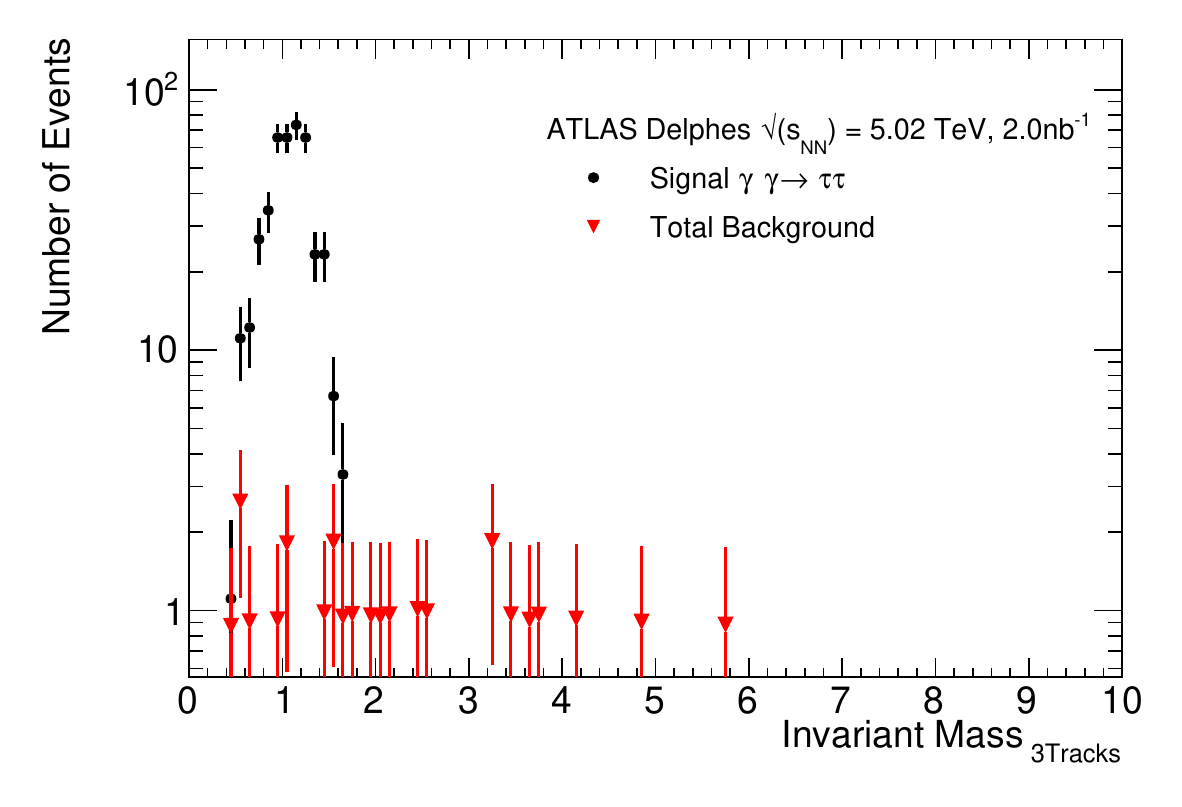}
  
 \caption{ 
    \label{fig:distribution_3trk} Distributions of the leading lepton ${p_{T}}$ and the invariant mass of the three tracks $\rm{Mass_{3T}}$ after the preselection for the signal region 1L3T.}
\end{figure}

The signal and background distributions of kinematic variables of interest after the preselection cuts are shown in Figure~\ref{fig:distribution} and~\ref{fig:distribution_3trk} for 1L1T and 1L3T signal region respectively. These distributions include all the background processes described in section~\ref{sec:bkg} and are normalised to 2.0 $\rm{nb^{-1}}$ of integrated luminosity.
The acoplanarity variable between the muon and the track (three tracks system) is defined as acoplanarity = $1-|\Delta\phi_{\mu,trk(s)}|/\pi$, while the missing transverse energy ($E_T^{Miss}$) is calculated from calorimeter energy deposits ($\vec{p}_T(i)$) as $ {E}_T^{Miss} = |\vec{E}_T^{Miss}|= |\sum_i \vec{p}_T(i)|$. 
The acoplanarity and the missing transverse energy distributions for 1L1T SR show, as expected, a strong difference between signal and background due to the presence of neutrinos from tau decays. 
The number of simulated background events, after the preselection for 1L3T SR, is very limited, however, the invariant mass of the three non-lepton tracks (Mass$\rm{_{3T}}$) plot shows a significant separation between signal and backgrounds.
The lepton $p_T$ for both signal regions do not show any significant discrimination between the signal and the background sample. The cut applied on muon $p_T$ is increased to 4 GeV for both the signal regions to apply the same efficiency of the electrons identification and to mimick the muon threshold used in the ATLAS trigger.  

The applied kinematic selection for the SC analysis is:
\begin{itemize}
    \item {\bf 1L1T}: 
   
    in order to reduce the overlap with the lepton, the track must fulfill an angular requirement: $\rm{\Delta R (lepton-trk) >0.02}$
    \footnote{$\Delta R$(lepton-trk)=$\sqrt{(\eta_{\mu,e} -\eta_{trk})^2
+(\phi_{lepton} - \phi_{trk})^2}$ where $\eta_{lepton},\phi_{lepton}$ and $\eta_{trk},\phi_{trk}$ are the pseudorapidities and the azimuthal angles of the lepton and of the track, respectively.}. 
The total charge of track plus lepton must be zero. In order to reduce the dilepton background, the lepton-track system is required to fulfill the cut: acoplanarity $<$ 0.4. 

    \item {\bf 1L3T}: 
   
    the three tracks are required not to overlap the lepton by applying the $\Delta R$ cut defined above on. The total charge of the three tracks plus the lepton must be zero.

The invariant mass of the three non-lepton tracks ($\rm{Mass_{3T}}$) is required to satisfy $\rm{Mass_{3T}}< 1.7$ to help the identification of the $\tau$ lepton. 
The acoplanarity$<$ 0.2 requirement is also applied to reduce the lepton background.

\end{itemize}

A summary of the SC selection for 1L1T and 1L3T is shown in Table~\ref{tab:cutflowing}.

\begin{table}[ht!]
\begin{center}
\begin{tabular}{|l|l|}
\hline
 \hline
Signal Region 1 Lepton &Signal Region 1 Lepton \\
 and 1 Track (SR1L1T) & and 3 Track (SR1L3T)\\
\hline
1 Lepton&1 Lepton \\
1 Track&3 Tracks \\
 $\rm{Charge_{1L1T}=0}$& $\rm{Charge_{1L3T}=0}$ \\
&$\rm{Mass_{3T}<}$1.7GeV \\
acoplanarity$<$0.4 &acoplanarity$<$0.2 \\
$p_T^{Muon}>$4GeV&${p_T^{Muon}>}$4GeV \\

    \hline
  \hline
\end{tabular}
\caption{ Selection cuts named as SC dedicated to the identification of the SRs applied to the lepton objects and to the tracks after the preselection cuts. 
\label{tab:cutflowing}}
\end{center}
\end{table}

\begin{figure}[ht!]
   \centering
    \includegraphics[width=0.49\linewidth]{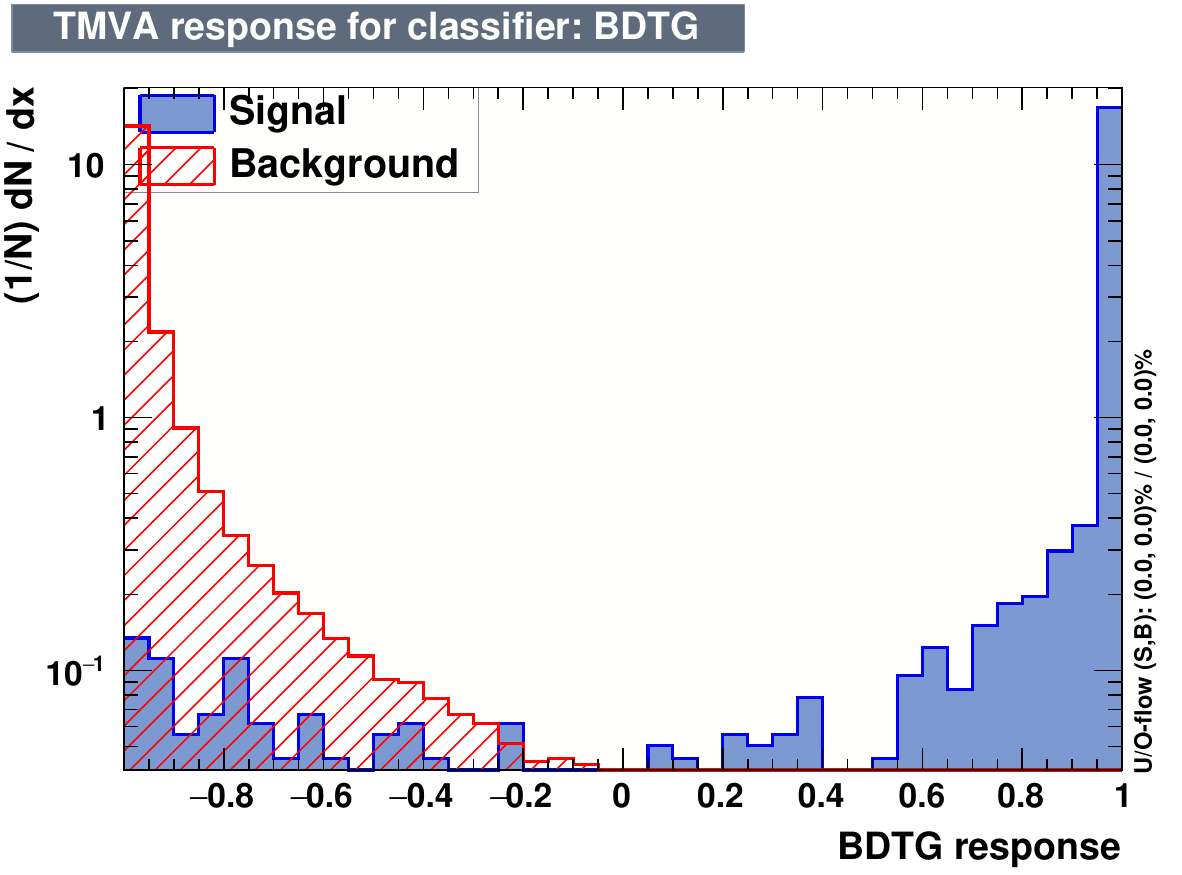}
    \includegraphics[width=0.49\linewidth]{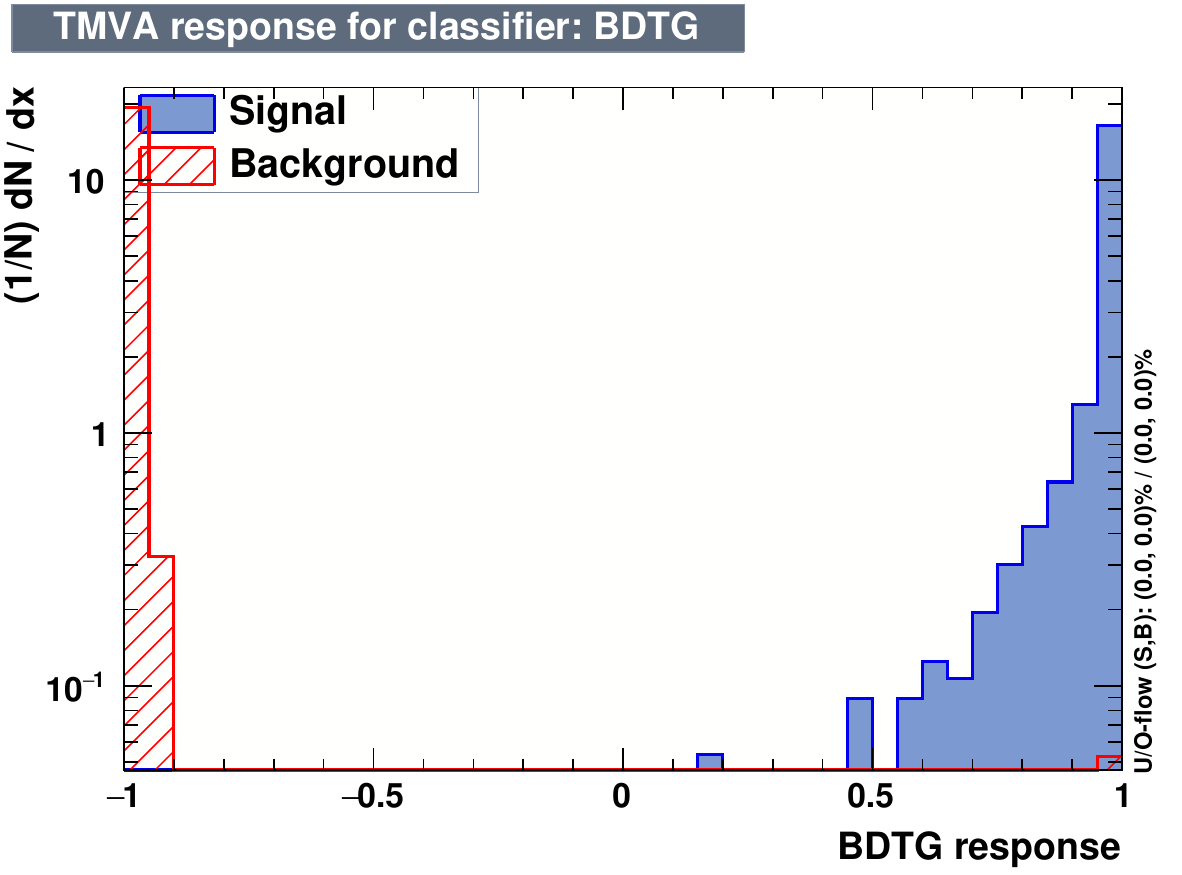}
 \caption{BDTG distributions for signal and background processes for the 1L1T (left) and 1L3T (right) signal regions.}
    \label{fig:BDTshapes}
\end{figure}

In order to investigate possible improvements in the signal over background ratio, a multivariate analysis has been implemented using 
gradient boosted decision tree (BDTG) in the TMVA framework~\cite{Hocker:2007ht}. The BDTG aims at improving the selection by fully exploiting the final state kinematical variables. The complete list of the variables used for the two signal regions, ordered by BDTG ranking, is reported in Table~\ref{tab:ranking}.  The BDTG distributions are shown in Figure~\ref{fig:BDTshapes} for signal and background processes for the two signal regions. 
The signal selection is obtained by applying thresholds on the BDTG distributions. The two thresholds, for 1L1T and 1L3T, are obtained based on best significance criterion with significance defined as $\rm{S/ \sqrt{S+B}}$. 

For 1L1T the BDTG threshold is set to  BDTG $>$ 0.84 corresponding to a significance  of 58 to be compared with a significance of 27 obtained with the SC analysis at 2 $\rm{nb^{-1}}$ of integrated luminosity.  For the signal region 1L3T the cut on BDTG is set to BDTG $>$ $-0.61$ with a significance of 20 to be compared with a significance of 18 obtained with the SC analysis at 2 $\rm{nb^{-1}}$ of integrated luminosity.

\begin{table}[ht!]
\begin{center}
\resizebox{9 cm}{!}{\begin{tabular}{||c|c||}
\hline
 {SR 1 Lepton and 1 Track } & {SR 1 Lepton and 3 Tracks }\\
  SR1L1T & SR1L3T\\
  \hline
 $\rm{\phi-Missing E_{T}}$& Sum $p_{T}$ 3 tracks\\
 Track $\eta$ & Invariant Mass\small{(Lepton+3Tracks)}\\
  Lepton $\phi$&  Lepton $P_{T}$\\
   Lepton $\eta$& Invariant Mass (3Tracks)\\
     Missing $E_{T}$ &$\Delta$R \small{(Lepton-3tracks~system)}\\
Acoplanarity&$\Delta \phi$ \small{(Lepton-3tracks~system)}\\
Track $P_{T}$&Missing $E_{T}$ \\
 Invariant Mass \small{(Lepton+Track)}&$\Delta$R (Lepton-Track)\\
  $\Delta$ R (Lepton-Track)&Track $P_{T}$\\
  Sum $P_{T}$ 3 tracks& $\Delta \phi$ (Lepton-track)\\
  Lepton $P_{T}$&Acoplanarity\\
  $\Delta \phi$ (Lepton-track)& $H_T$ ($\sum_i |\vec{p}_T(i)|$)\\
$H_T $ ($\sum_i |\vec{p}_T(i)|$)& \\

   \hline
\end{tabular}}
\caption{ The BDTG ranking of the variables used, divided per SR.
\label{tab:ranking}}
\end{center}
\end{table}

\section{Sensitivity to the tau anomalous magnetic moment  }

In this section the sensitivity to the signal strength $\mu_{\tau\tau}$, defined as the ratio of the observed signal yield to the SM expectation assuming the SM value for $\mu_{\tau\tau}$ =1, and to the anomalous magnetic moment \atau, are presented. Both estimates, carried out by using a profile-likelihood fit ~\cite{TRexFitter} on the lepton transverse momentum distributions, are obtained for the SC and the BDTG analyses.

\begin{figure}[ht!]
   \centering
   \includegraphics[width=0.49\linewidth]{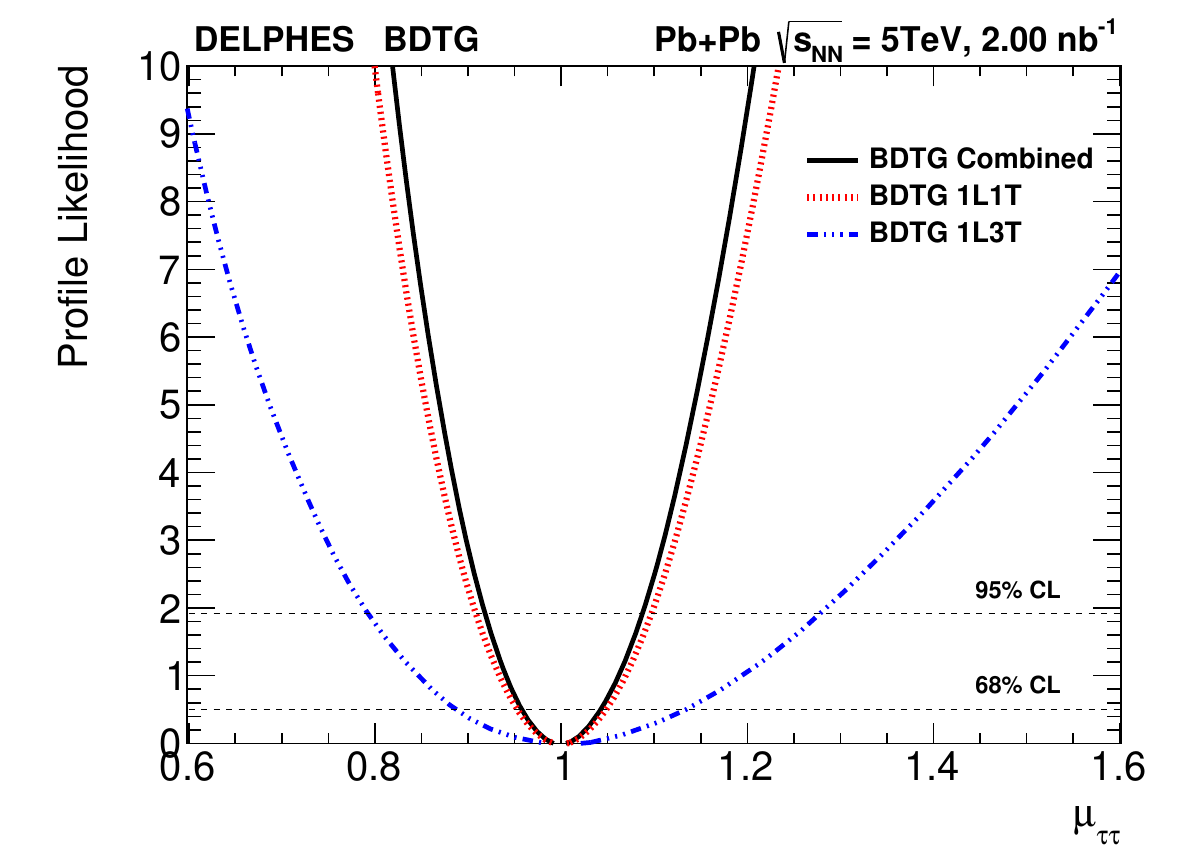}
    \includegraphics[width=0.49\linewidth]{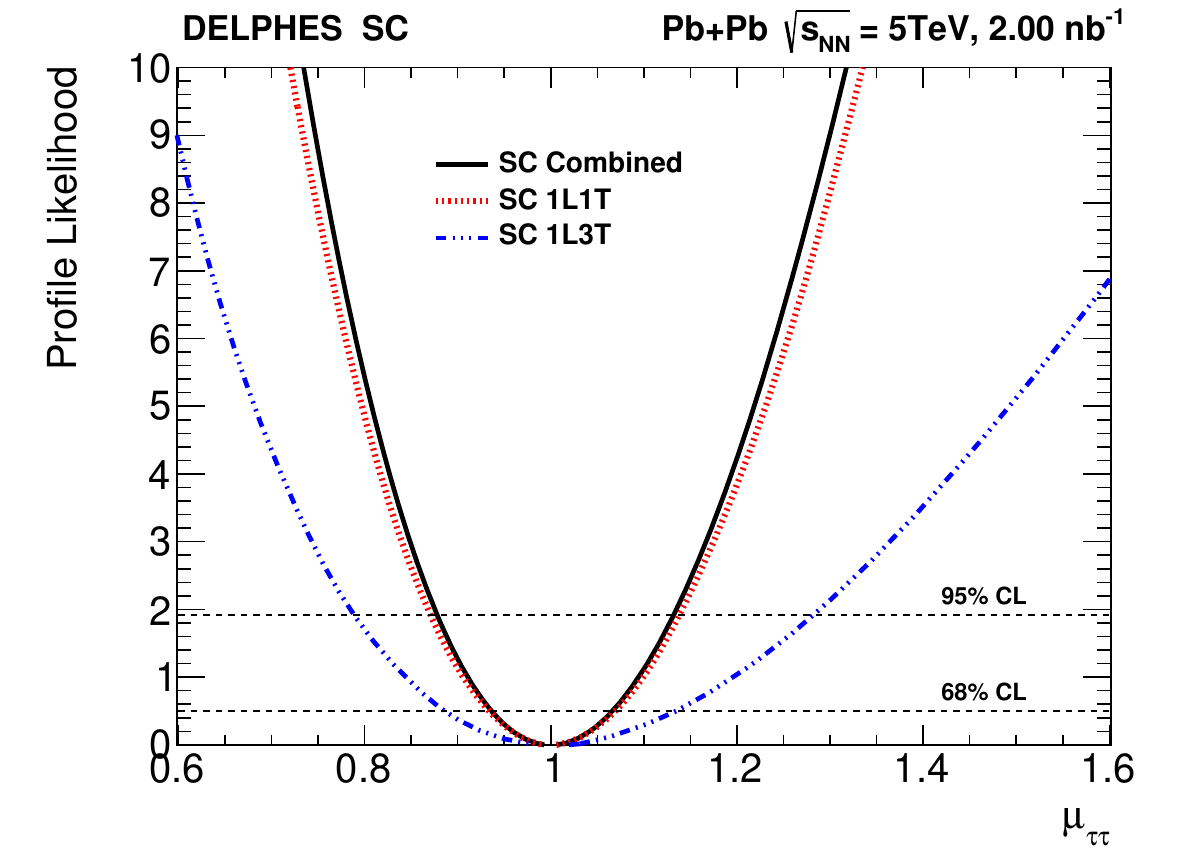}
 \caption{Profile likelihood scan of the signal strength parameter using Asimov Data and considering \atau=0, for the two signal regions. The left plot shows the results for the BDTG selection while the right plot for the SC. The normalization systematic uncertainties included are: 2\% to mimic the ATLAS luminosity uncertainties and an additional 10$\%$ to overall mimic experimental conditions. The integrated luminosity is set to 2.0 $nb^{-1}$. \label{fig:mu2}}
\end{figure}

\begin{figure}[ht!]
  \centering
      \includegraphics[width=0.7\linewidth]{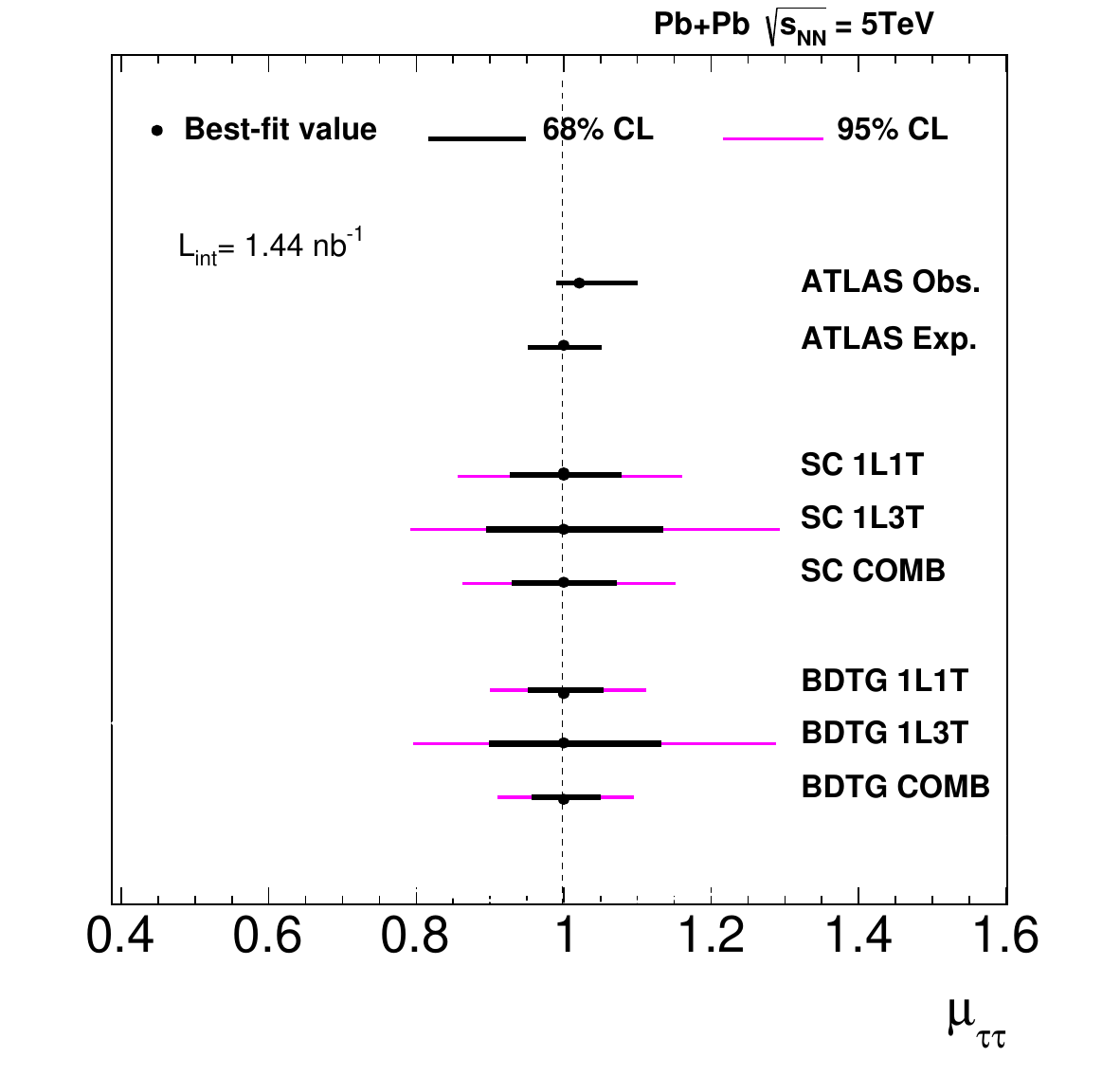}

 \caption{ Sensitivity for $\mu_{\tau\tau}$ signal strength using Asimov Data for the two signal regions and for the combination using the BDTG and SC selections using 1.44 $\rm{nb^{-1}}$ of integrated luminosity. The normalization systematic uncertainties included are the ATLAS luminosity estimated as 2$\%$ and an additional 10$\%$ to overall mimic experimental conditions.
 These results are compared with existing results from ATLAS (expected and observed) obtained by using 1.44 $\rm{nb^{-1}}$ of integrated luminosity\cite{ATLASg-2}. A point denotes the best-fit value for each measurement where available, while thick black (thin magenta) lines show 68\% CL (95\% CL) intervals.}
    \label{fig:sum_mu}
\end{figure}

The sensitivity to the signal strength $\mu_{\tau\tau}$ at CL 95\%  are measured to be ${\mu_{\tau\tau}= 1^{-0.121}_{+0.130}}$ and $\rm{{\mu_{\tau\tau}= 1^{-0.078}_{+0.084}}}$ for the SC and BDTG analysis respectively.
This estimates are obtained using Asimov Data.
The normalization systematic uncertainties included are: the luminosity estimated to be 2\% and an additional 10\% to take into account the experimental conditions. 
These results are illustrated in the plots of Figure~\ref{fig:mu2} where a clear improvement of the $\mu_{\tau\tau}$ precision is shown with the BDTG approach. The sensitivity obtained for the two signal regions and for the two analysis selections presented in this work are compared to the ATLAS measurement in Figure~\ref{fig:sum_mu}. In this Figure~\ref{fig:sum_mu}, the integrated luminosity of the two analysis selections is scaled to 1.44 $nb^{-1}$, the same integrated luminosity of the quoted ATLAS results.

\begin{table}[ht!]
\begin{center}
\begin{tabular}{||c|c|c|c||}
\hline
95\% CL& {SR 1L1T } & {SR 1L3T } & {Combined} \Tstrut\Bstrut\\ 
\hline \hline \Tstrut\Bstrut
SC & ${\mu_{\tau\tau}}$ = 1 $_{+0.137} ^{-0.127}$ &${\mu_{\tau\tau}}$ = 1 $_{+0.280} ^{-0.199}$& ${\mu_{\tau\tau}}$ = 1 $_{+0.130} ^{-0.121}$\\[0.1cm]
BDTG &${\mu_{\tau\tau}}$ = 1 $_{+0.080} ^{-0.083}$ &${\mu_{\tau\tau}}$ = 1 $_{+0.277} ^{-0.195}$& ${\mu_{\tau\tau}}$ = 1 $_{+0.084} ^{-0.078}$\\[0.1cm]
  \hline
 \hline \Tstrut\Bstrut
 SC & ${a_{\tau}}$ = 0 $_{+0.023} ^{-0.030}$ &${a_{\tau}}$ = 0 $_{+0.026} ^{-0.035}$& ${a_{\tau}}$ = 0 $_{+0.021} ^{-0.026}$\\[0.1cm]
BDTG & ${a_{\tau}}$ = 0 $_{+0.012} ^{-0.020}$ &${a_{\tau}}$ = 0 $_{+0.022} ^{-0.038}$& ${a_{\tau}}$ = 0 $_{+0.011} ^{-0.019}$\\ [0.1cm]
   \hline\hline
\end{tabular}
\caption{The sensitivity to $\mu_{\tau\tau}$ and $a_{\tau}$ at 95\% CL for each signal region and the combination of the two. The two methods: BDTG and SC are compared. The integrated luminosity is set to 2.0 $nb^{-1}$.
\label{tab:mu_results}}
\end{center}
\end{table}

\begin{figure}[htbp]
    \includegraphics[width=0.48\linewidth]{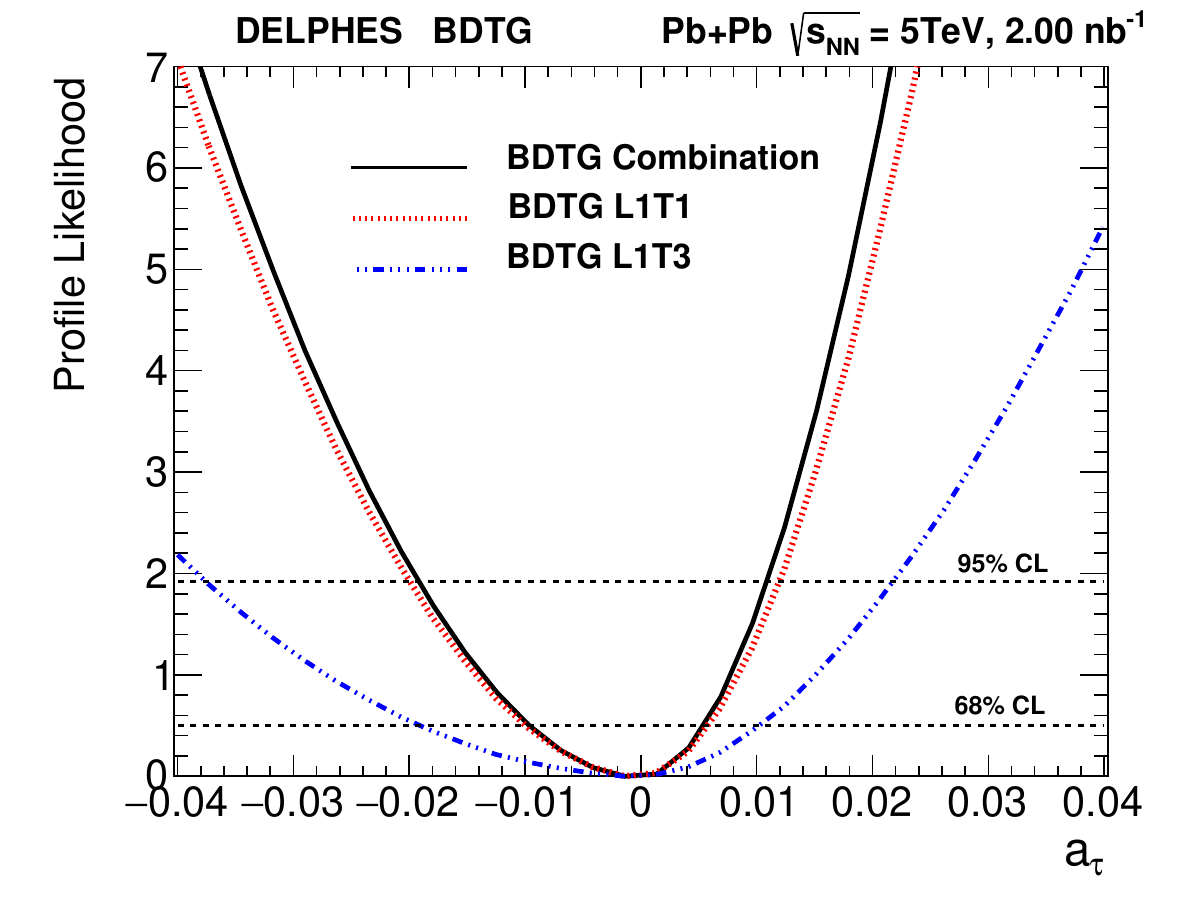}
    \includegraphics[width=0.50\linewidth]{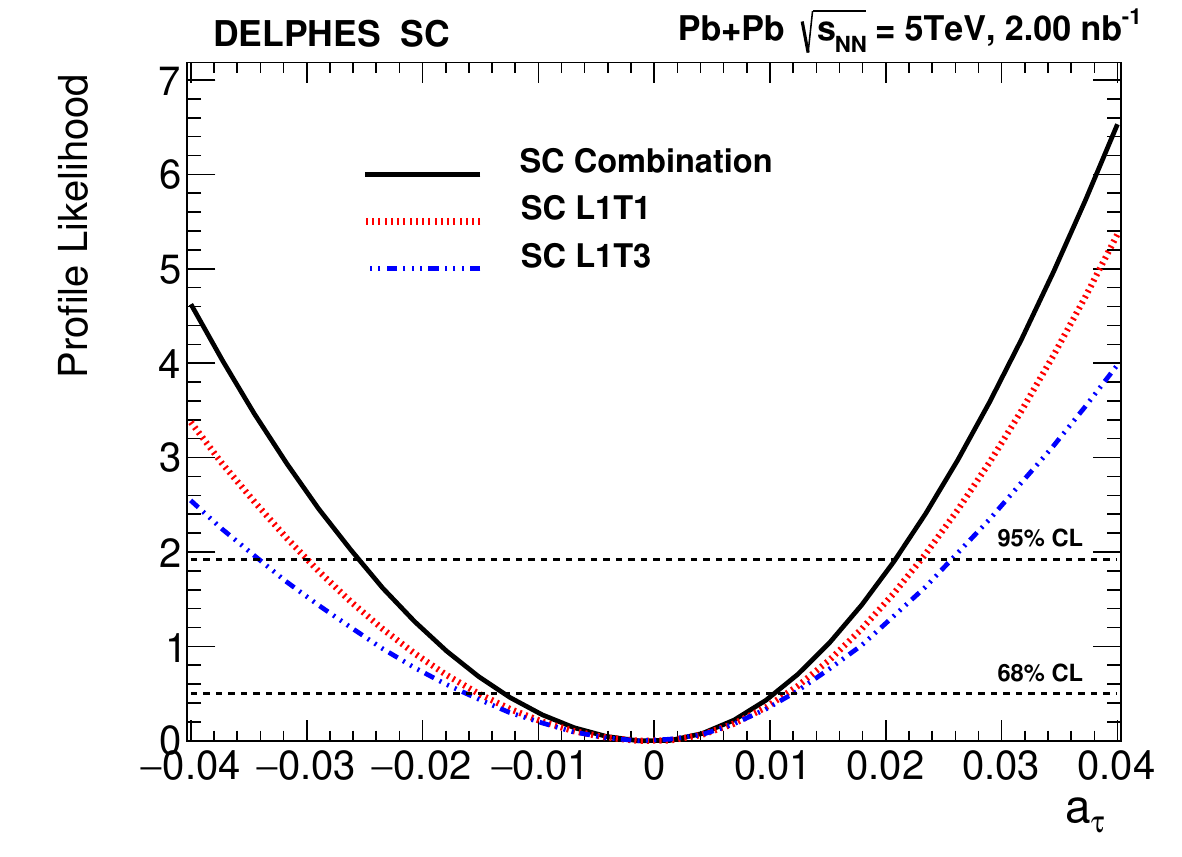}
 \caption{Profile likelihood for \atau using Asimov Data for the two signal regions and the combination of the two regions. The left plot shows the results for the BDTG selection while the right plot for the SC. The normalization systematic uncertainties included are: 2\% to mimic the ATLAS luminosity uncertainties and an additional 10$\%$ to overall mimic experimental conditions. The integrated luminosity is set to 2.0 $nb^{-1}$.}
    \label{fig:atau1}
\end{figure}

The sensitivity to \atau is estimated with a fit where \atau is the only free parameter and using the lepton transverse momentum distribution with a nominal value of \atau set to the SM value (\atau=0). Simulated signal samples with various \atau values are included in the fit.
The profile likelihood scan are presented in Figure~\ref{fig:atau1}. The sensitivity to $a_{\tau}$ at 95\% CL are $\rm{a_{\tau}}$ = ${0^{-0.19}_{+0.11}}$  and  $\rm{a_{\tau}}$ = 0 $_{+0.021} ^{-0.026}$ using the BDTG and SC analysis respectively. A clear improvement in the sensitivity to \atau is shown when using the BDTG approach. 

\begin{figure}[ht!]
  \centering
      \includegraphics[width=0.8\linewidth]{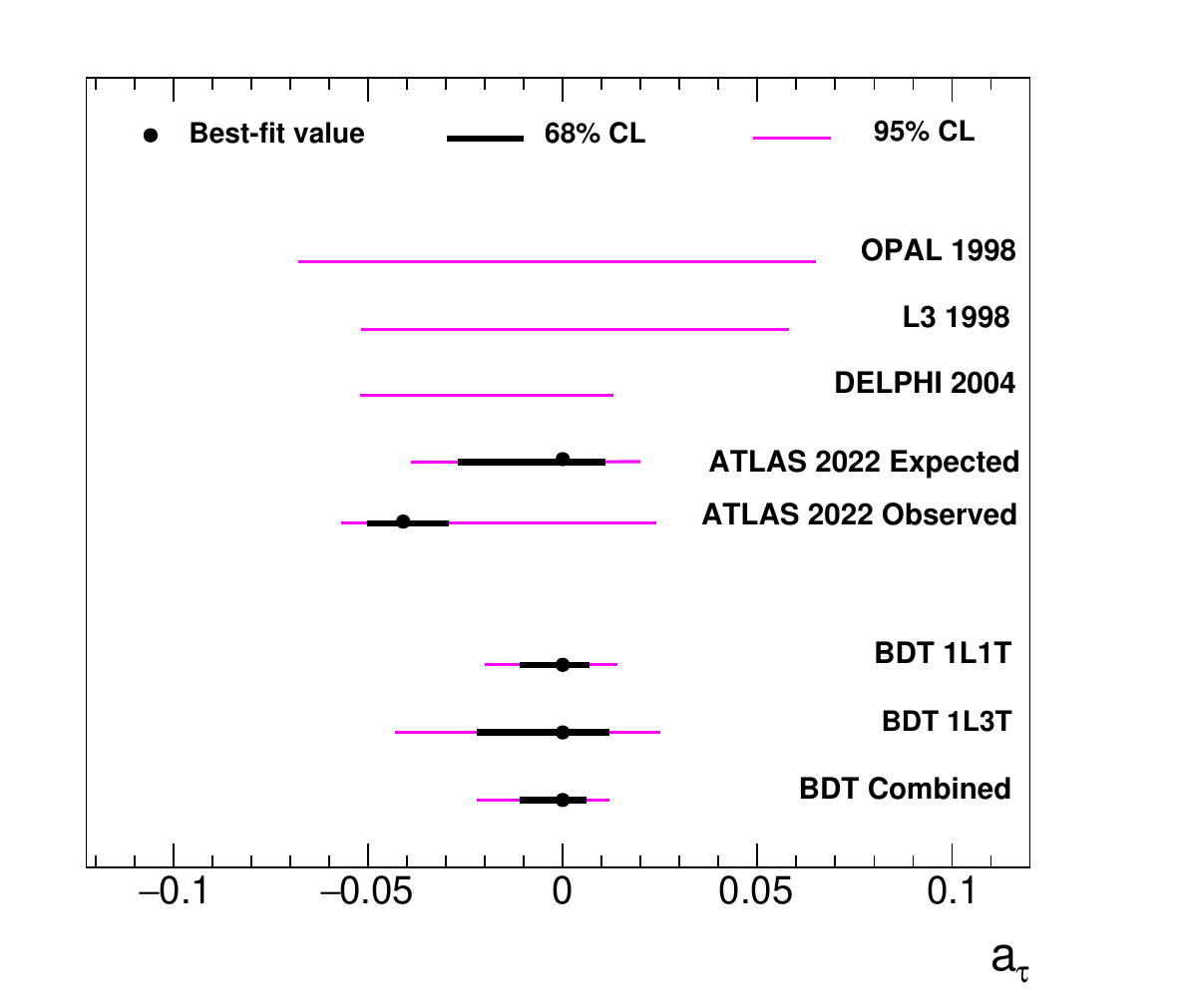}
 \caption{ Best-fit value of \atau parameter using Asimov Data for the two signal regions and the combination using the BDTG selection with an integrated luminosity of 1.44 $\rm{nb^{-1}}$. The systematic uncertainties included are the ATLAS luminosity estimated as 2$\%$ and an additional 10$\%$ to overall mimic experimental conditions. These results are compared with existing results from OPAL\cite{OPAL}, L3\cite{L3}, DELPHI\cite{DELPHI:2003nah} and the latest results from ATLAS obtained with an integrated luminosity of 1.44 $\rm{nb^{-1}}$ (expected and observed limits). A point denotes the best-fit value for each measurement where available, while thick black (thin magenta) lines show 68\% CL (95\% CL) intervals.}
    \label{fig:atau2}
\end{figure}

The sensitivity obtained on \atau with the BDTG analysis is compared with previous measurements in Figure~\ref{fig:atau2}. The integrated luminosity of the BDTG analysis in this figure is scaled to 1.44 $nb^{-1}$, the same integrated luminosity of the latest ATLAS results.

\clearpage

\section{Conclusions}
\paragraph{}
A study was  presented  of the ultra-peripheral process $Pb(\gamma)- Pb(\gamma) \rightarrow \tau^+ \tau^-$ using a signal and background simulation,  based on experimental conditions and detector performances of the ATLAS experiment at the CERN LHC. One $\tau$ is required to decay leptonically while the other one  decays hadronically, into one or three tracks. This study  aims at an estimation of the precision in  the signal strength $\mu_{\tau\tau}$ measurement and  of the sensitivity achievable in the determination of the $\tau$ anomalous magnetic moment $a_{\tau}$.
\par  A different approach than in previous studies was adopted by using the signal events produced by an effective $a_\tau$-generating Lagrangian term, implemented in the MadGraph5 Monte Carlo generator. Signal and background events were normalised at a luminosity of $\rm{2.0\, {nb^{-1}}}$. The signal selection was performed with a SC procedure and with a new BDTG approach. 
\par  As a result:
\begin{itemize}
    \item 
the signal strength $\rm{{\mu_{\tau\tau}= 1^{-0.078}_{+0.084}}}$ at 95\% CL  was achieved with the BDTG selection, to be compared with   ${\mu_{\tau\tau}= 1^{-0.127}_{+0.130}}$ obtained with the SC procedure. 
    \item 
The sensitivity to $a_{\tau}$ at 95\% CL resulted to be $\rm{{\mu_{\tau\tau}= 0^{-0.019}_{+0.011}}}$ by  using the BDTG method and $\rm{a_{\tau}}$ = 0 $_{+0.021} ^{-0.026}$ by using the SC selection. 
\end{itemize}

\par Our results show that, using the BDTG approach, a significant improvement in precision  could be obtained  for both  $\mu_{\tau\tau}$ and $a_{\tau}$  determinations compared  to the latest limits published by the ATLAS experiment \cite{ATLASg-2}. 
The present expected ATLAS sensitivity to $a_\tau$ is about 0.06 (at 95 \%CL)
dominated by statistics (0.045).
\par In this work we introduced a systematic
uncertainty of 2\% from LHC luminosity and an additional  conservative
systematic uncertainty of 10\% on the overall $\gamma\gamma \rightarrow \mu\mu$ production cross section yielding a sensitivity to $a_\tau$ of 0.03
and 0.047 for the BDT approach and  the cut-flow procedure respectively, with
an improvement of $\approx 60$ \% in favour of BDT. 

\par However  these results look still insufficient  to explore new physics. It would be desirable to obtain at least  a sensitivity  of $ \approx  10^{-3}$ in $ a_\tau$ measurement, so approaching the order of magnitude  expected  in the SM,  dominated by  1 loop contribution in QED. 
\par In fact it is worth to note that there are new physics models
predicting $a_\tau$ as large as $\approx 10^{-3}$ (\cite{Feruglio:2018fxo}).
\par
We believe that  with the upcoming experiments in proton-proton and heavy ions collisions at LHC and $e^+e^-$ collisions at Belle2, thanks to higher collected luminosities, but also with the help of new  analysis procedures, the  $a_\tau$ measurements  could be performed with a precision better  by at least one order of magnitude  than done in the past, providing a new window in search for new physics.

\section*{Acknowledgements}
We thank g-2 tau ATLAS working group, in particular M. Dyndal, L. Beresford, for the useful discussion on the expected $a_{\tau}$ results and private communications.
We thank P. Paradisi and A. Strumia for useful discussions.
NV thanks G. Landini, A. Strumia and J. Wang for help in the validation of the UFO model.

\newpage
\newpage
\clearpage
\appendix

\section{Monte Carlo distributions and Cross Sections \label{MonteCarlo}}

Several background samples have been included in this analysis, two millions  events for each sample have been generated and simulated. Table \ref{tab::samples} reports the complete list of the samples used with the production cross section and the expected number of events at 2.0 $\rm{nb^{-1}}$ of integrated luminosity.
Cuts on lepton $p_{T} >1$ GeV and ${\eta}<2.5$ are applied at generation level.
\begin{table}[ht!]
\begin{center}
 \resizebox{8cm}{!}{\begin{tabular}{|l|c|c|}
       \hline
       Sample  & Cross Section ($pb$) & events@2$\rm{nb^{-1}}$\\
       \hline
       SM ($a_{\tau}=0$)  &  $5.49 \times10^8 \pm 1.7 \times10^5 $ &1111111\\
       SM+BSM ($a_{\tau}=+0.02$) & $5.79 \times10^8 \pm 1.9 \times10^5 $&1176470 \\
              SM+BSM ($a_{\tau}=-0.02$) & $5.22 \times10^8 \pm 1.8 \times10^5 $ &1052631\\
                 SM+BSM ($a_{\tau}=-0.01$) & $5.35 \times10^8\pm 1.7\times 10^5 $&1081081 \\
                    SM+BSM ($a_{\tau}=+0.01$) & $5.64 \times10^8\pm 1.8\times 10^5$ &1142857\\
                       SM+BSM ($a_{\tau}=-0.04$) & $4.99 \times10^8 \pm1.6 \times10^5 $ &998000\\
                        SM+BSM ($a_{\tau}=+0.04$) & $6.12 \times10^8 \pm1.9 \times10^5 $ &1212121\\
        $\gamma\gamma\rightarrow e^{-}e^{+}$ & $4.258\times 10^8 \pm 1.8\times 10^8$&869565\\
 $\gamma\gamma\rightarrow \mu^{-}\mu^{+}$   &$4.258 \times10^8 \pm 1.8 \times10^8$ &869565\\
  $\gamma \gamma\rightarrow bb$  &$1.629\times 10^6 \pm  2,3 \times10^2 $&3257\\
   $\gamma\gamma\rightarrow cc$   &$3.276 \times10^6\pm 1.3\times 10^5$ &6557\\
    $\gamma\gamma\rightarrow jet(c,d,u) jet(c,d,u)$   &$3.686 \times10^6 \pm  1.5 \times10^5$& 7380\\

       \hline
    \end{tabular}}
\end{center}
\caption{Total cross sections of each sample included in the analysis. A cut on $p_{T} >1$ GeV and ${\eta}<2.5$ of the lepton is applied at generation level. Different signal samples have been produced depending on the anomalous magnetic coupling value. } \label{tab::samples}
\end{table}

\begin{figure}[ht!]
    \centering
    \includegraphics[width=0.48\linewidth]{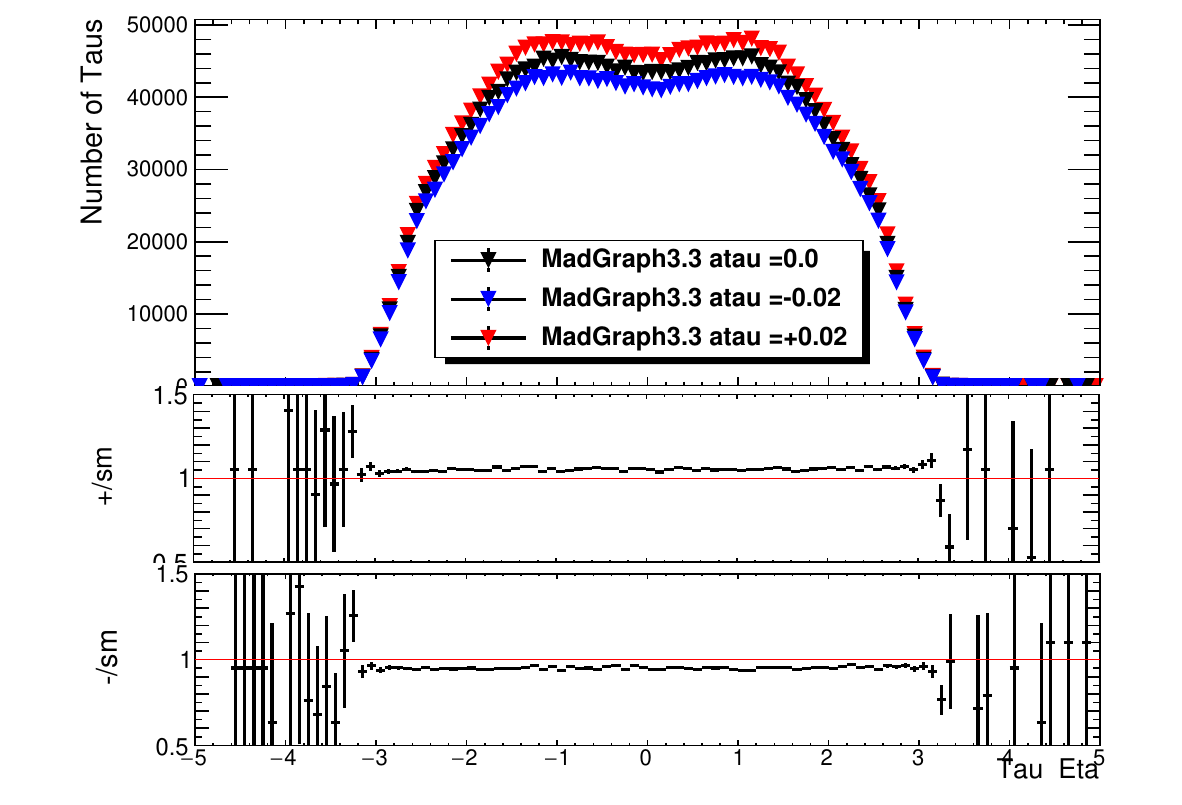}
    \includegraphics[width=0.48\linewidth]{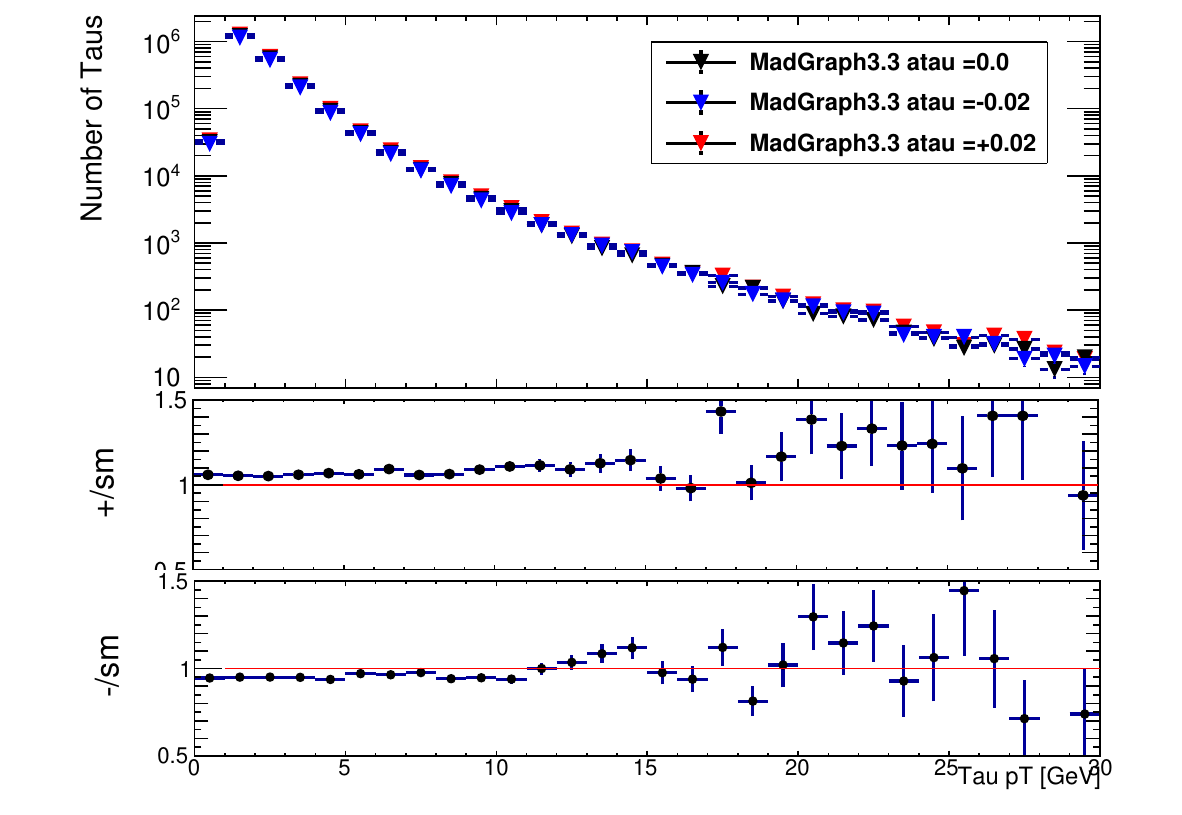}
   \includegraphics[width=0.48\linewidth]{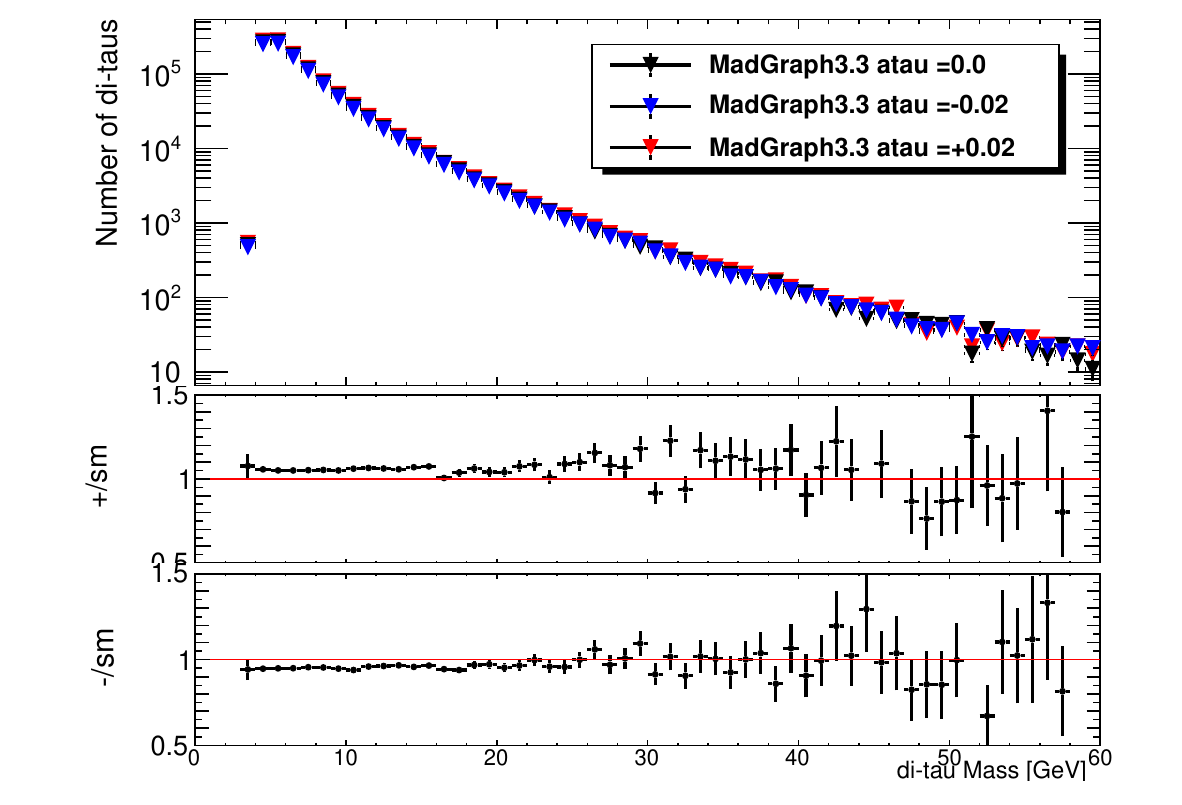}
    \includegraphics[width=0.48\linewidth]{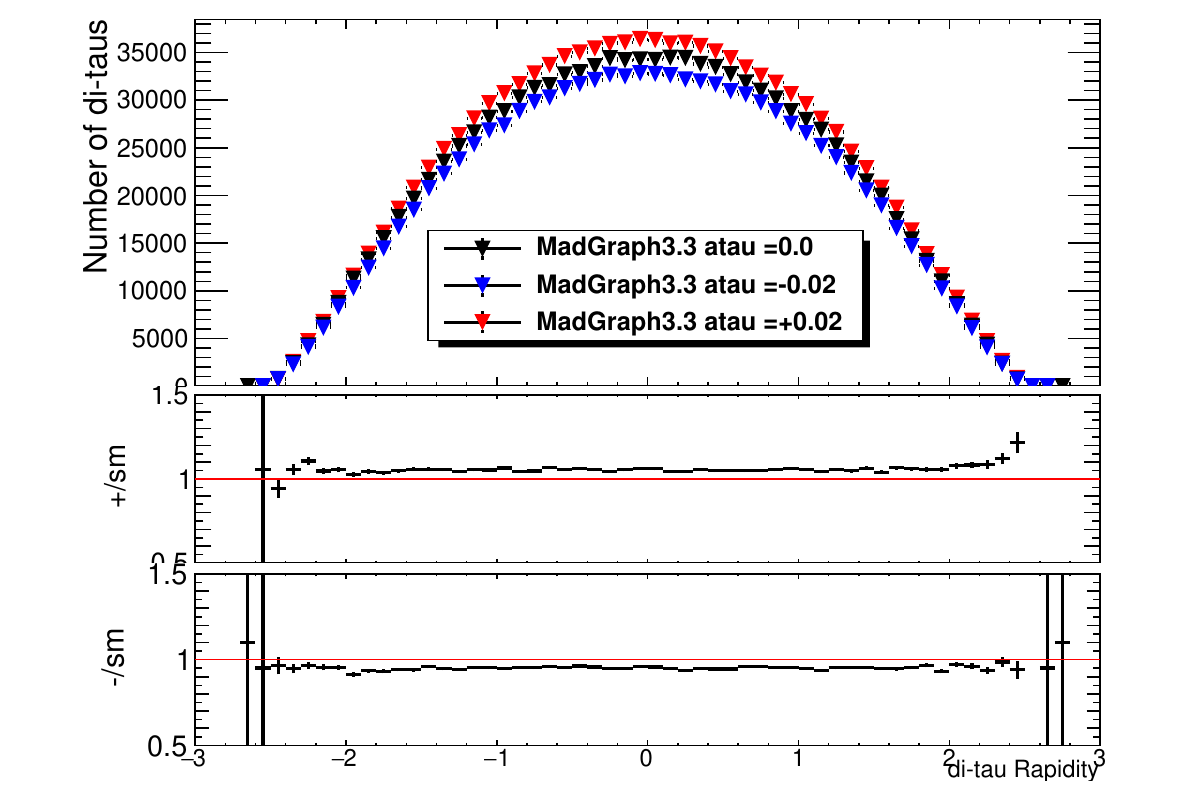}
    
    \caption{Top plots: Tau $\eta$ and $p_T$ distributions at different values of $a_{tau}$: +0.02, -0.02, 0. The ratio between $a_{\tau}$ =$\pm$0.02 and $a_{\tau}$=0 is reported in the bottom side of each plots. Bottom plots:  di-tau system mass and rapidity distributions at different values of $a_{\tau}$: +0.02, -0.02, 0. The ratios are reported in the bottom side of the plots. }
    \label{fig:at_002}
\end{figure}

Figure~\ref{fig:at_002} shows the distributions of tau and di-tau system for the value of $\rm{a_{\tau} = \pm 0.02}$ compared with the nominal Standard Model $a_{\tau}=0$.

\newpage

\section{Boosted Decision Tree}
In this appendix, the BDTG observables used in the evaluation are shown for each channel. The background sample used is the sum of the background channels already shown in Table~\ref{tab::samples}.

\begin{figure}[htp]
    \centering
    {\includegraphics[width=\linewidth]{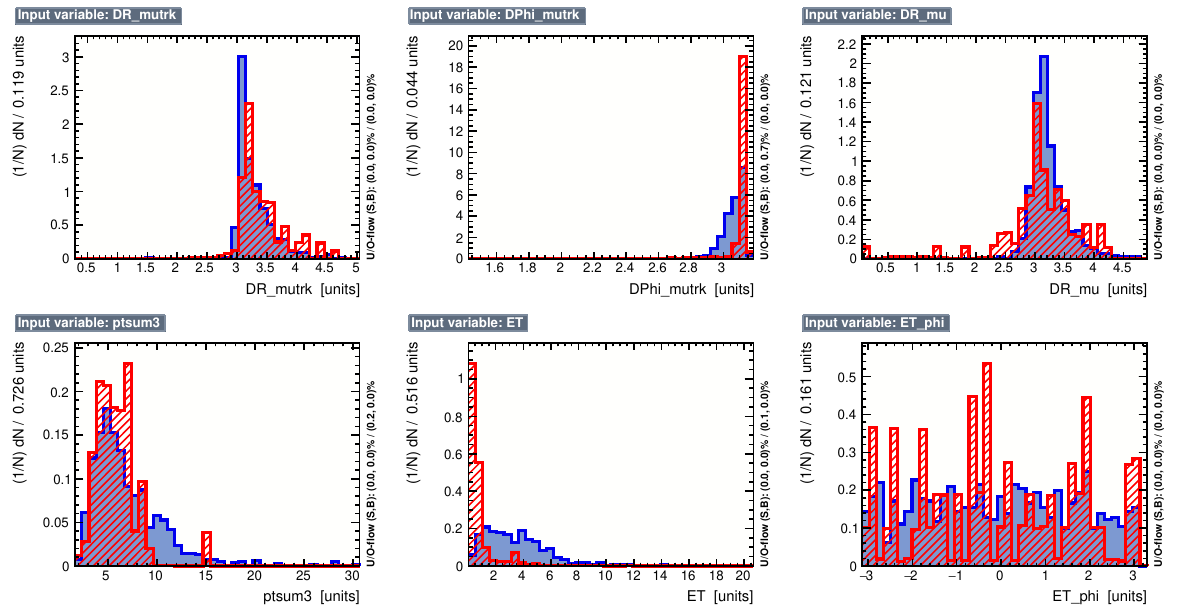}}
    {\includegraphics[width=1\linewidth]{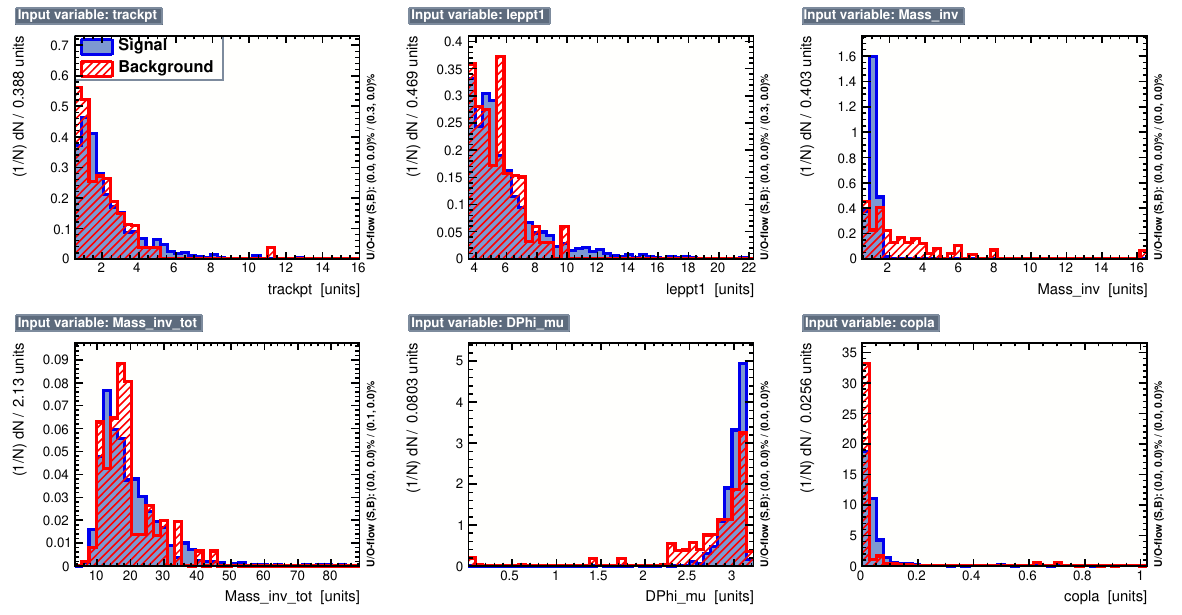}}
     
    \caption{Distributions of the observable used for the SR1L3T BDT analysis.
 }
    \label{fig::BDT2}
\end{figure}

\begin{figure}[ht!]
    \centering
    
    {\includegraphics[width= 1\linewidth]{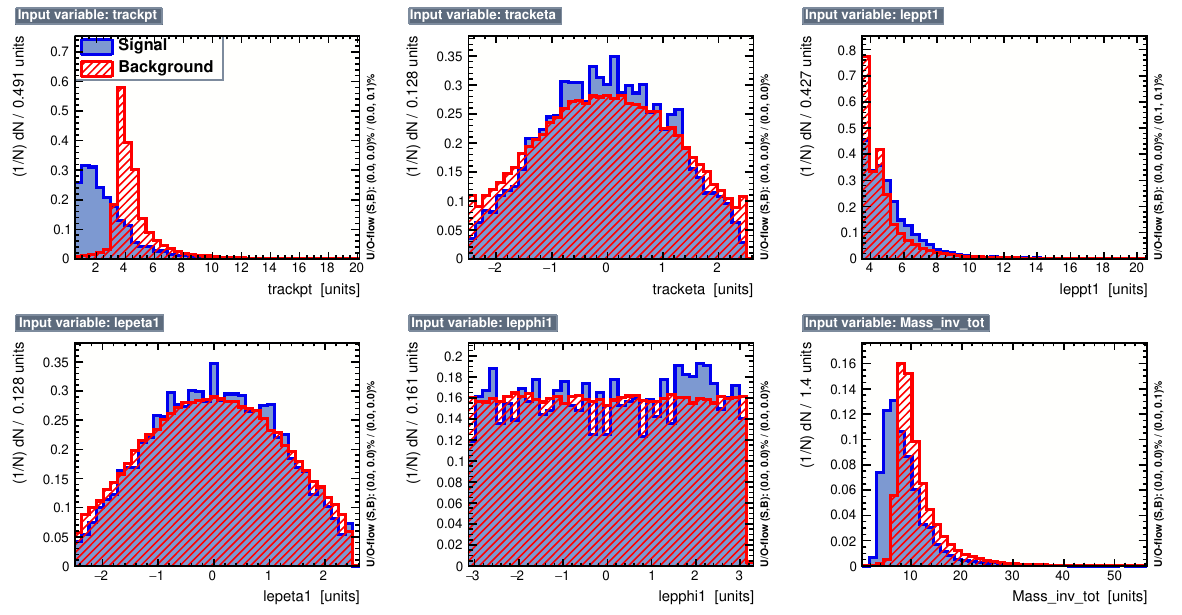}}\\
        
    {\includegraphics[width= 1\linewidth]{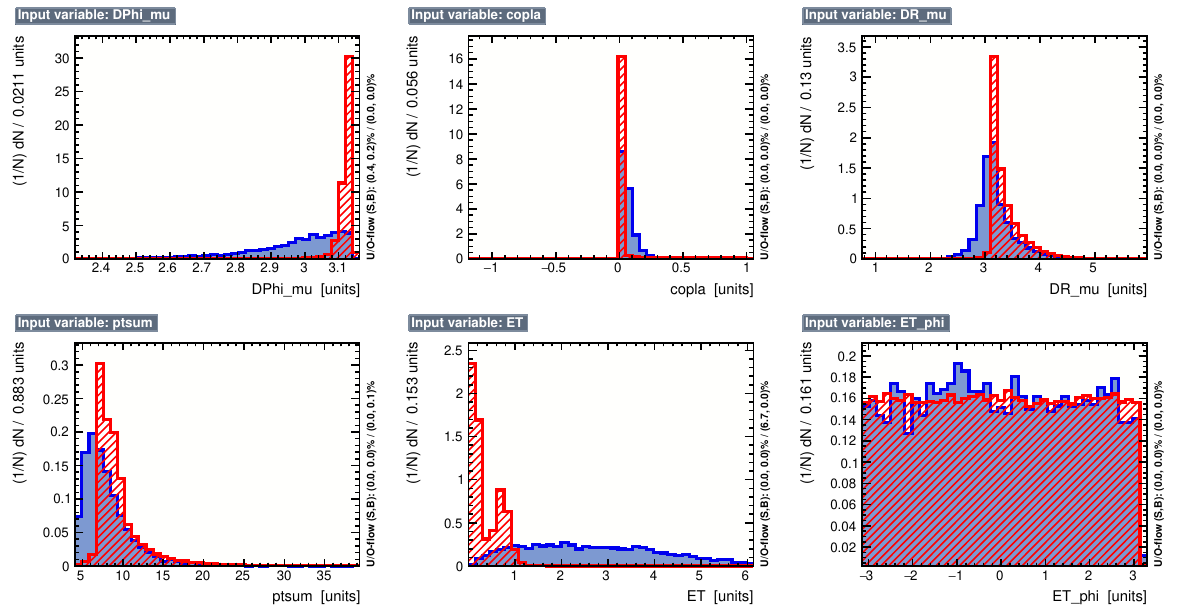}}\\
    

    \caption{ Observables used for the SR1L1T BDT analysis.
 }
    \label{fig::BDT1}
\end{figure}

The Figure~\ref{fig::BDT1} and the Figure~\ref{fig::BDT2} report for 1L1T and 1L3T respectively the distributions of the  variables used for the selection.

\section{Profile Likelihood at 1.44 nb-1 integrated luminosity.\label{app_cuts}}

In this appendix, the profile likelihoods obtained for 1.44 $\rm{nb^{-1}}$  integrated luminosity are shown. This integrated luminosity corresponds to the integrated luminosity collected with heavy ions collisions in the year 2018 (2015 excluded).

\begin{figure}[ht!]
\centering
   \includegraphics[width=0.48\linewidth]{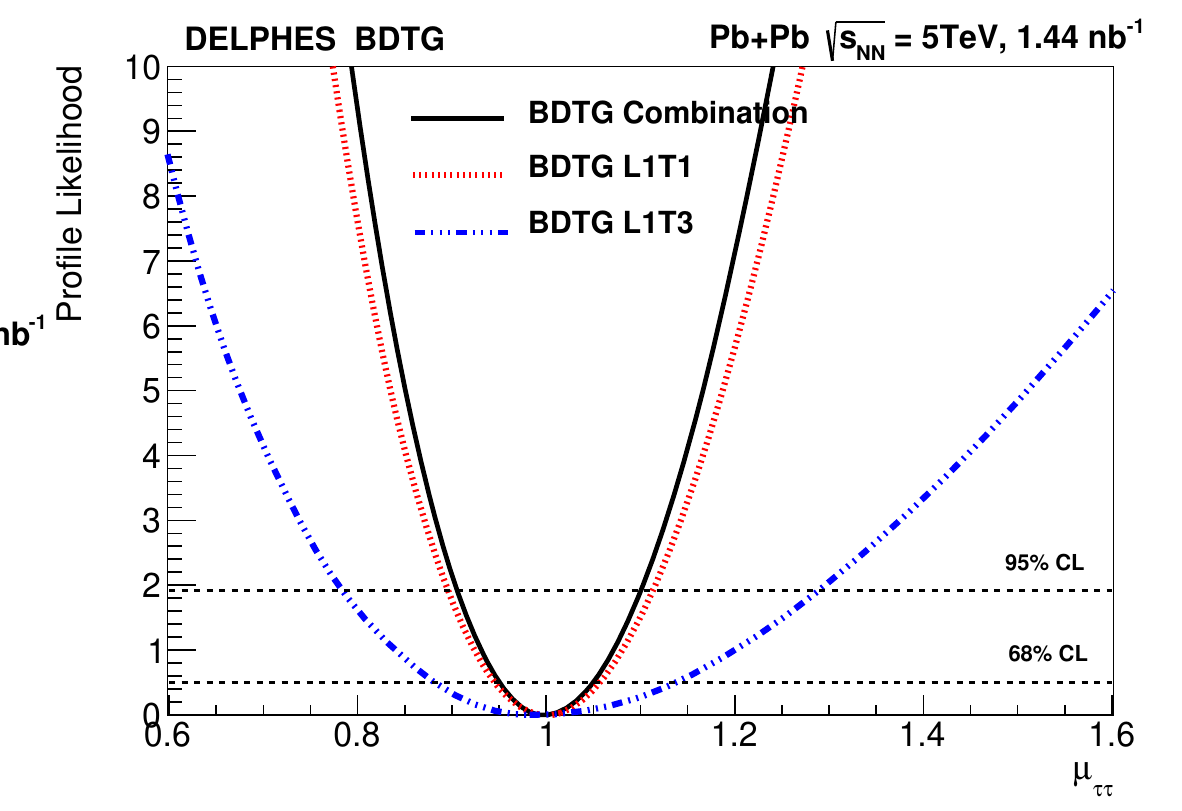}
    \includegraphics[width=0.48\linewidth]{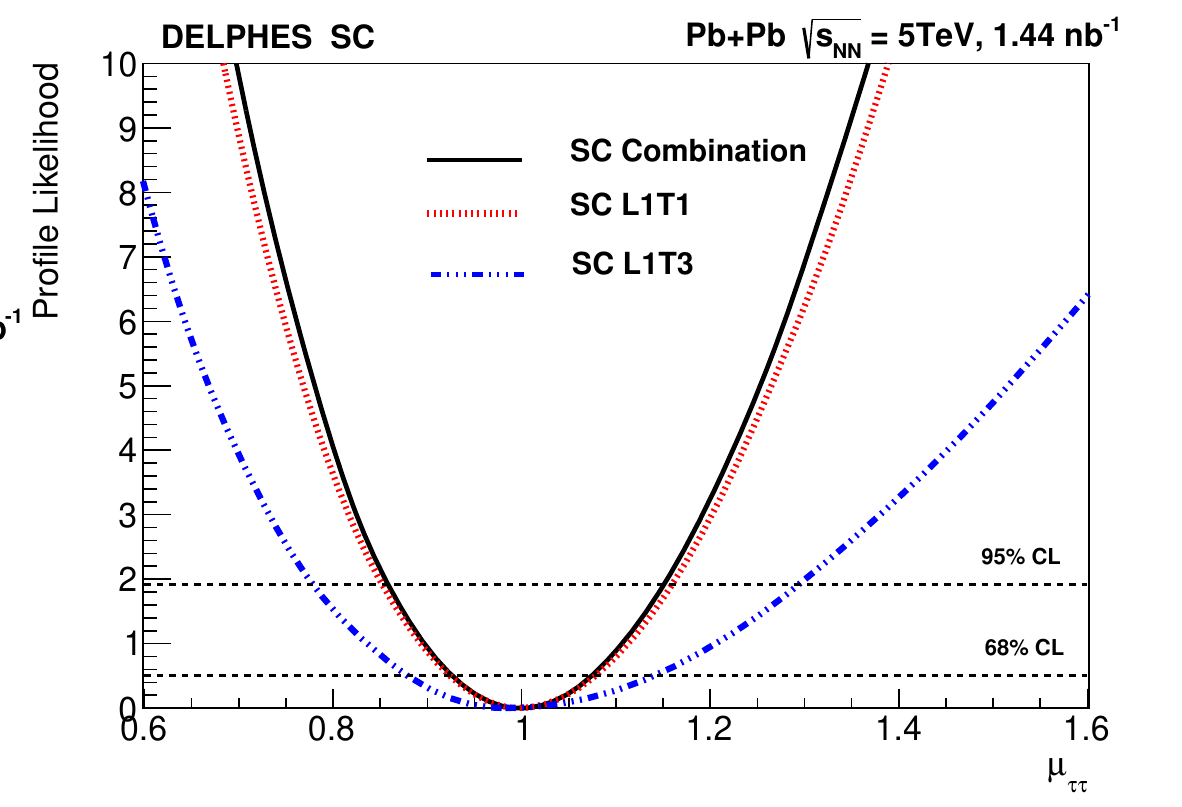}
 \caption{Profile likelihood scan of the signal strength parameter using Asimov Data and considering \atau=0, for the two signal regions. The left plot shows the results for the BDTG selection while the right plot for the SC. The systematic uncertainties included are on the luminosity, estimated to be 1.44$\%$ and an additional 10$\%$ to conservatively mimic the experimental conditions.
    \label{fig:mu2_2}}
\end{figure}

The sensitivity to the signal strength $\mu_{\tau\tau}$ at CL 95\%  are measured to be ${\mu_{\tau\tau}= 1^{-0.137}_{+0.152}}$ and $\rm{{\mu_{\tau\tau}= 1^{-0.09}_{+0.095}}}$ for the SC and BDTG analysis respectively. This estimates are obtained using Asimov Data.
The normalization systematic uncertainties included are: the luminosity estimated to be 2\% and an additional 10\% for a more realistic estimate of the experimental conditions. 
These results are illustrated in the plots of Figure~\ref{fig:mu2_2} where a clear improvement of the $\mu_{\tau\tau}$ precision is shown with the BDTG approach. 

\begin{table}[ht!]
\begin{center}
 \resizebox{8cm}{!}{\begin{tabular}{||c|c|c|c||}
\hline
& {SR 1L1T } & {SR 1L3T } & {Combined} \Tstrut\Bstrut\\ 
\hline \hline \Tstrut\Bstrut

 SC (95\% CL)& ${\mu_{\tau\tau}}$ = 1 $_{+0.161} ^{-0.144}$ &${\mu_{\tau\tau}}$ = 1 $_{+0.293} ^{-0.208}$& ${\mu_{\tau\tau}}$ = 1 $_{+0.152} ^{-0.137}$\\[0.1cm]
BDTG (95\% CL)&${\mu_{\tau\tau}}$ = 1 $_{+0.112} ^{-0.100}$ &${\mu_{\tau\tau}}$ = 1 $_{+0.288} ^{-0.204}$& ${\mu_{\tau\tau}}$ = 1 $_{+0.095} ^{-0.09}$\\[0.1cm]
  \hline
 \hline \Tstrut\Bstrut
 SC (95\% CL)& ${a_{\tau}}$ = 0 $_{+0.027} ^{-0.034}$ &${a_{\tau}}$ = 0 $_{+0.029} ^{-0.040}$& ${a_{\tau}}$ = 0 $_{+0.024} ^{-0.029}$\\[0.1cm]
BDTG (95\% CL)& ${a_{\tau}}$ = 0 $_{+0.014} ^{-0.022}$ &${a_{\tau}}$ = 0 $_{+0.025} ^{-0.043}$& ${a_{\tau}}$ = 0 $_{+0.012} ^{-0.022}$\\ [0.1cm]
   \hline\hline

\end{tabular}}
\caption{The sensitivity to $\mu_{\tau\tau}$ and $a_{\tau}$ at 95\% CL for each signal region and the combination of the two. The two methods: BDTG and SC are compared.
\label{tab:mu_results_2}}
\end{center}
\end{table}

\begin{figure}[htbp]
  \includegraphics[width=0.49\linewidth]{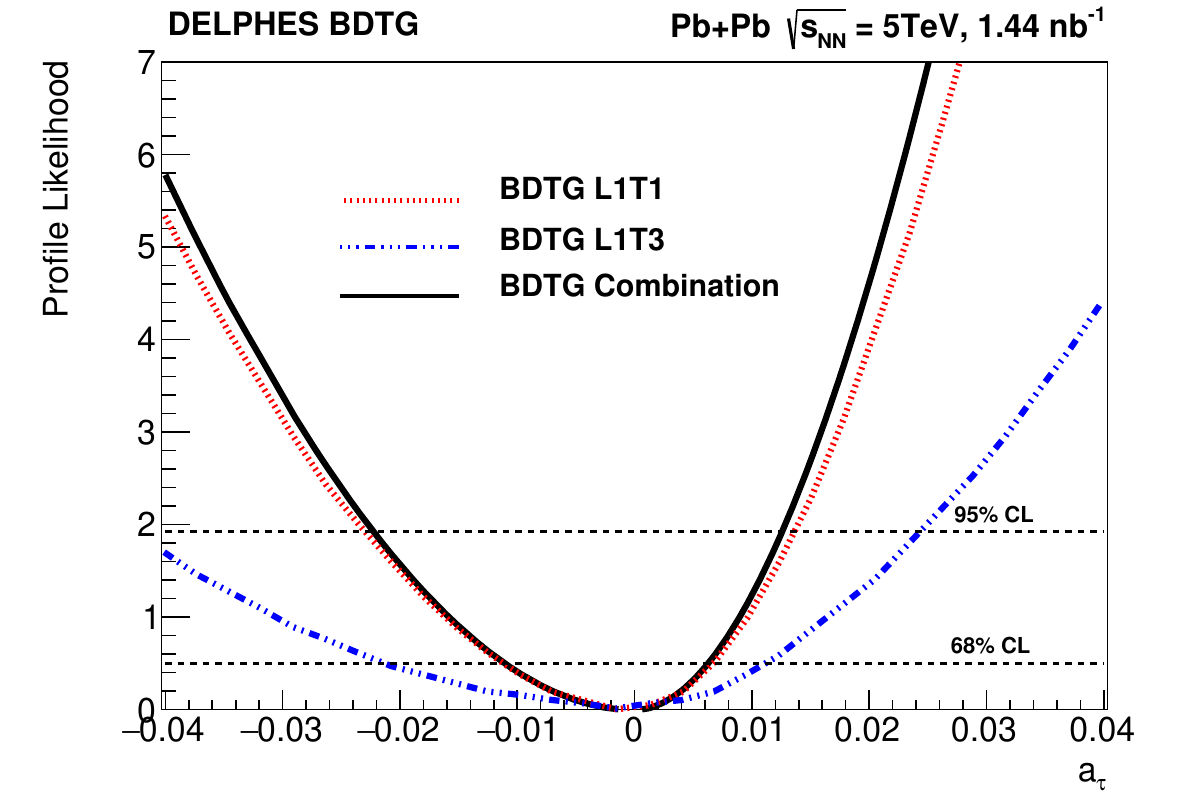}
    \includegraphics[width=0.49\linewidth]{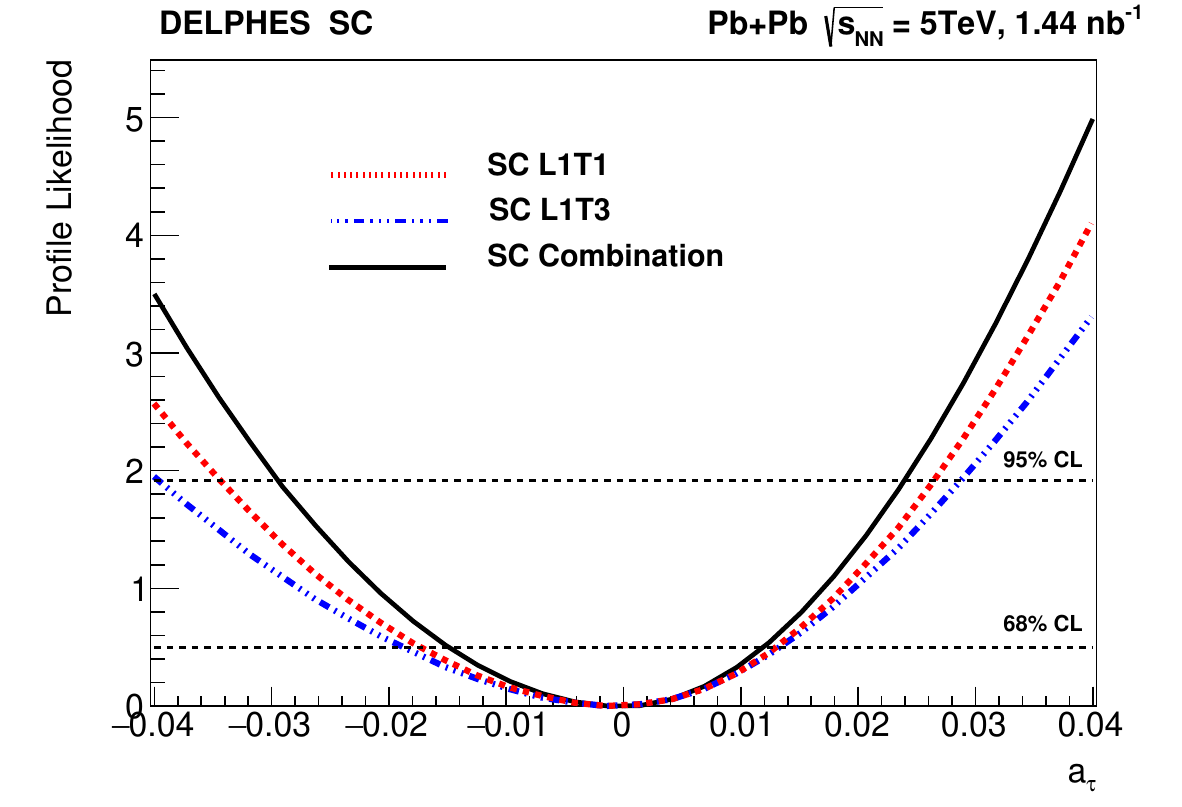}
  \caption{Profile likelihood for \atau using Asimov Data for the two signal regions and the combination of the two regions. The left plot shows the results for the BDTG selection while the right plot for the SC. The systematic uncertainties included are: 2\% to mimic the ATLAS luminosity uncertainties and an additional 10$\%$ to overall mimic experimental conditions.}
    \label{fig:atau1_2}
\end{figure}

The sensitivity to \atau is estimated with a fit where \atau is the only free parameter and using the lepton transverse momentum distribution with a nominal value of \atau set to the SM value (\atau=0). Simulated signal samples with various \atau values are included in the fit.
The profile likelihood scan are presented in Figure~\ref{fig:atau1_2}. The sensitivity to $a_{\tau}$ at 95\% CL are $\rm{a_{\tau}}$ = 0 $_{+0.012} ^{-0.022}$ and  $\rm{a_{\tau}}$ = 0 $_{+0.024} ^{-0.029}$ using the BDTG and SC analysis respectively. A clear improvement in the sensitivity to \atau is shown when using the BDTG approach. 

\begin{figure}[ht!]
  \centering
    \includegraphics[width=0.8\linewidth]{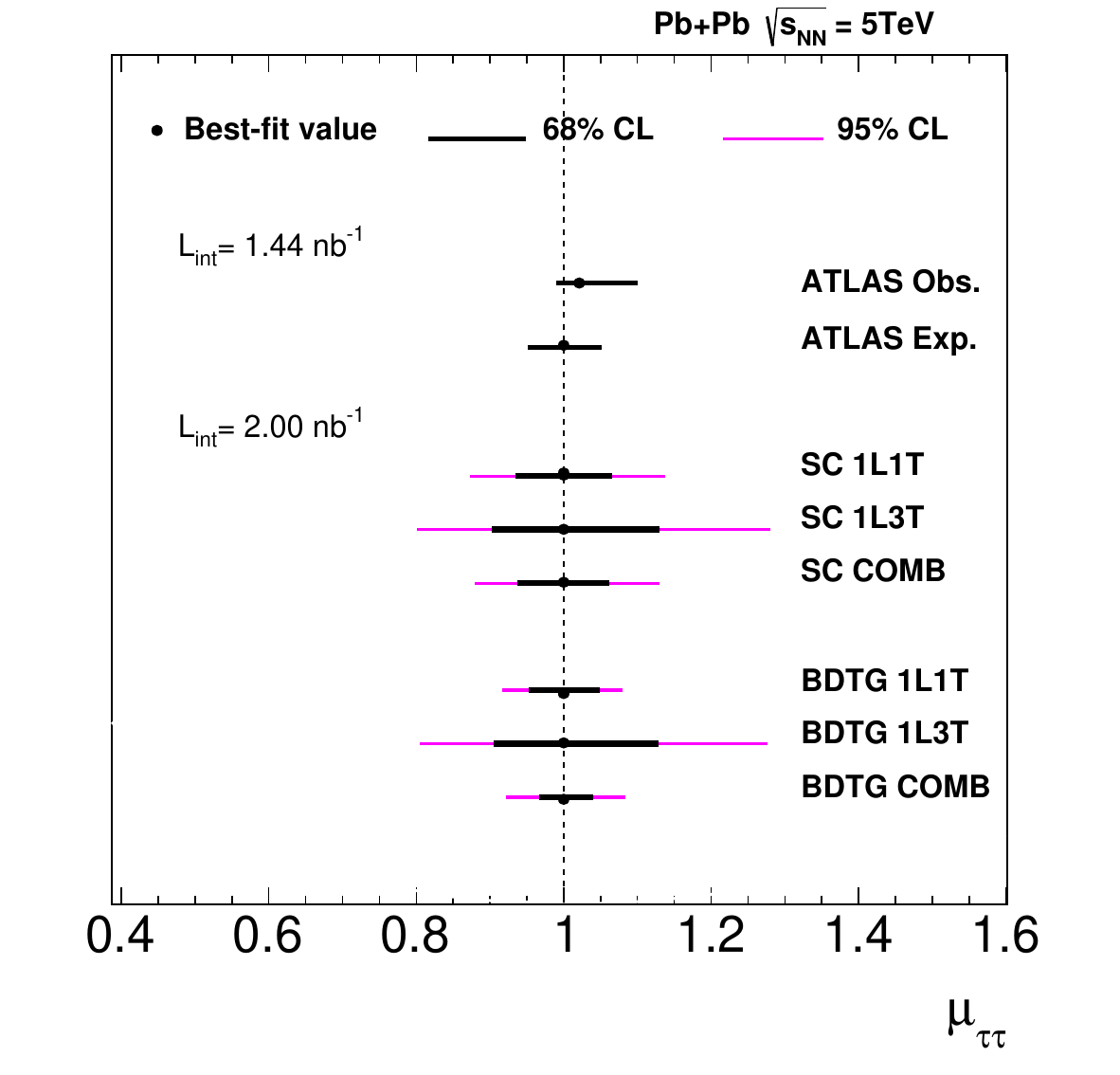}

 \caption{ Sensitivity for $\mu_{\tau\tau}$ signal strength using Asimov Data for the two signal regions and for the combination using the BDTG and SC selections using 2.0 $\rm{nb^{-1}}$ of integrated luminosity. The systematic uncertainties included are the ATLAS luminosity estimated as 2$\%$ and an additional 10$\%$ to overall mimic experimental conditions. These results are compared with existing results from ATLAS (expected and observed) obtained by using 1.44 $\rm{nb^{-1}}$ of integrated luminosity\cite{ATLASg-2}. A point denotes the best-fit value for each measurement where available, while thick black (thin magenta) lines show 68\% CL (95\% CL) intervals.}
    \label{fig:sum_mu_2}
\end{figure}

\begin{figure}[ht!]
  \centering
      \includegraphics[width=0.8\linewidth]{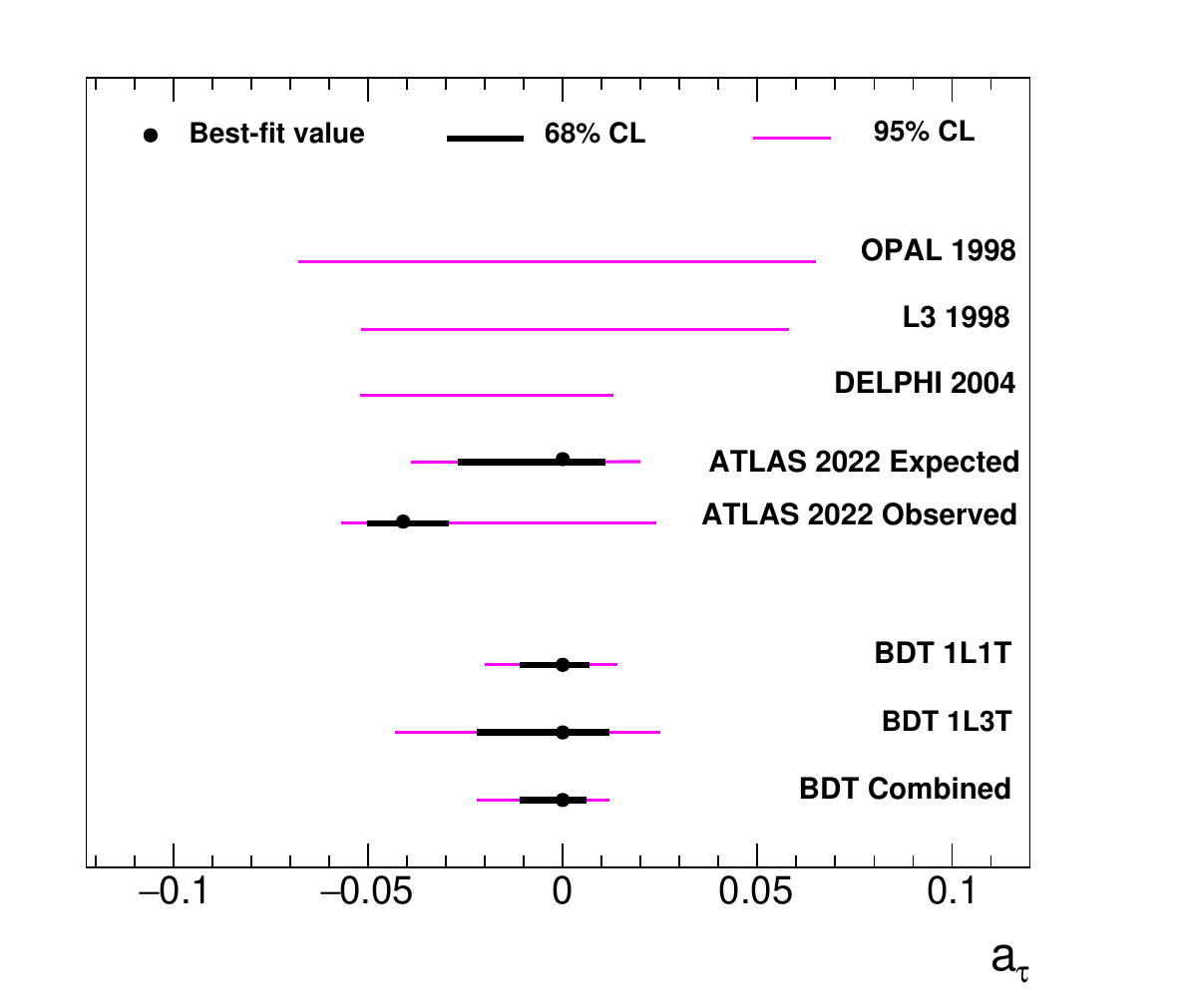}
 \caption{ Best-fit value of \atau parameter using Asimov Data for the two signal regions and the combination using the BDTG selection with an integrated luminosity of 2.0 $\rm{nb^{-1}}$. The systematic uncertainties included are the ATLAS luminosity estimated as 2$\%$ and an additional 10$\%$ to overall mimic experimental conditions. These results are compared with existing results from OPAL\cite{OPAL}, L3\cite{L3}, DELPHI\cite{DELPHI:2003nah} and the latest results from ATLAS obtained with an integrated luminosity of 1.44 $\rm{nb^{-1}}$ (expected and observed limits). A point denotes the best-fit value for each measurement where available, while thick black (thin magenta) lines show 68\% CL (95\% CL) intervals.}
    \label{fig:atau2_2}
\end{figure}

The sensitivity obtained on \atau with the BDTG analysis and with an integrated luminosity of 2.0 $\rm{nb^{-1}}$ is compared with previous measurements in Figure~\ref{fig:atau2_2}. 

\section{Selection Cut Flow \label{cuts}}

In this appendix, the selection cuts applied to all the signal regions are shown. The di-lepton signal region is also included for completeness, however the low statistics preclude the test of the BDT method. Therefore, the two leptons signal region (2LSR) is also removed from the final limits comparison.
The Table~\ref{tab:cutflowbis} and Table~\ref{table:2lep} report the event yields after each cut normalised to 2 $\rm{nb^{-1}}$ of integrated luminosity for the 1L1TSR, 1L3TSR and 2LSR.

\begin{table*}[ht!]
\begin{center}
\small\small
\begin{tabular}{l|c|c|c|c|c|c|c}
  \hline
  \hline
\hline
 \hline
Selection&\atau&\atau&\atau&\atau&\atau&\atau&\atau\\
  Cuts  & -0.04& -0.02 & -0.01 
    &SM 0& +0.01& +0.02 & +0.04\\
    \hline
    \hline
    
Total Event & 1000000&1052631&1081081&1111111&1142857&1176470&1212121\\\hline 
\multicolumn{8}{l}{Signal Region 1 Lepton and 1 Track (SR1L1T)}\\ \hline
\hline
\hline
1 Lepton &  5828.5& 5804.41	&5766.66&6075.59&6113.13&6580.31&	7057.2\\ \hline
1 Track & 3917&	3905.02	&3923.1	&4031.52&4091.79&4422.94&4804.2 \\\hline
 $Charge_{1L1T}=0$ &3853&3845.06&3864.24&3967.14&4030.69&4349.44&4729.2\\\hline
Acoplanarity$<$0.4&1757.5&1811.02&1773.9&1893.11&1872.31&2029.78&2149.2\\\hline
$P_T^{Muon}>$4GeV &1320&1336.04&1318.14&1403.04&1374.4&1513.51&1596.6\\\hline
$E_T^{Miss}>$1GeV & 1220.5&1237.15&1213.92&1283.16&1259.63&1392.97&1480.2\\\hline
 \hline
    \hline
\multicolumn{8}{l}{Signal Region 1 Lepton and 3 Track (SR1L3T)}\\
 \hline
    \hline
1 Lepton & 5828.5&	5804.41&	5766.66&	6075.59&	6113.13&	6580.31&	7057.2\\\hline
3 Tracks & 422&	410.28&	371.52&	416.81	&433.39	&450.99	&488.4\\\hline
$Charge_{1L3T}=0$ & 420.5	&409.23&	369.36&	416.25&	431.68&	450.41&	487.2\\\hline
$Mass_{3T}<$1.7GeV & 420&	403.97&	365.58&	413.48&	426.54&	449.23&	484.8\\\hline
Acoplanarity$<$0.2 &403	&383.98&	345.06&	390.72&	403.13&	420.42&	459.6 \\\hline
$P_T^{Muon}$>4GeV &344&	327.70&	299.7	&323.01	&336.32	&355.74	&397.8 \\\hline
  \hline
\end{tabular}
\caption{Event yield after each cut at 2\rm{$nb^{-1}$} for each \atau value generated. \label{tab:cutflow}}
\end{center}
\end{table*}

\begin{table*}[ht!]
\begin{center}
\small\small
\begin{tabular}{l|c|c|c|c|c|c}
 \hline
  \hline
     Selection  & $\gamma \gamma\rightarrow \tau \tau$ & $\gamma \gamma\rightarrow \mu\mu $&$\gamma \gamma\rightarrow ee$ &$\gamma \gamma\rightarrow bb $&$\gamma \gamma\rightarrow cc$& $\gamma \gamma\rightarrow jj $ \\
\hline
Total Event &1111111 &869565 &869565 &3245.91 &6557.38 & 7380.07\\ \hline
\multicolumn{7}{l}{Signal Region 1 Lepton and 1 Track (SR1L1T)}\\ \hline
1 Lepton & 6081.06& 57964.6& 35241.6& 18.96& 0.31&0.03 \\
\hline
1 Track &4035.15 &54400.6 &27396.2 &1.43 &0.05 & 0\\\hline
 $Charge_{1L1T}=0$ &3970.71 &54399.7 &27395 &0.88 &0.02 & 0\\\hline
Acoplanarity$<$0.4& 1894.81&1193.53 &435.71 & 0.52& 0.003& 0 \\\hline
$P_T^{Muon}>$4GeV &1404.3 &746.75& 435.71& 0.31& 0.003& 0\\\hline
 \hline
    \hline
\multicolumn{7}{l}{Signal Region 1 Lepton and 3 Track (SR1L3T)}\\
 \hline
    \hline
1 Lepton &6081.06& 57964.6& 35241.6&18.96 &0.31 &0.03 \\\hline
3 Tracks &417.18 & 13.62& 5.53& 3.82&0.09 &0.09 \\ \hline
$Charge_{1L3T}=0$ &416.63 &13.19 &5.53 &1.91 & 0.05& 0\\ \hline
$Mass_{3T}<$1.7GeV &413.85 &5.96 & 2.55& 0.40& 0.01&0.01 \\ \hline
Acoplanarity$<$0.2 &391.07 & 5.96&1.70 &0.35 &0.01 &0.01 \\ \hline
$P_T^{Muon}>$4GeV &323.30 &4.68 &1.70 &0.23 &0.01 &0.01\\ \hline
    \hline
\end{tabular}
\caption{Event yield after each cut at 2\rm{$nb^{-1}$} for 1L1TSR and 1L3TSR} .  
\label{tab:cutflowbis}
\end{center}

\end{table*}

\begin{table*}[ht!]
\begin{center}
\small\small
\begin{tabular}{l|c|c|c|c|c|c}

    \hline
    \hline
    Selection  & $\gamma \gamma\rightarrow \tau \tau$ & $\gamma \gamma\rightarrow \mu\mu $&$\gamma \gamma\rightarrow ee$ &$\gamma \gamma\rightarrow bb $&$\gamma \gamma\rightarrow cc$& $\gamma \gamma\rightarrow jj $ \\
\hline
Total Event &1111111 &869565 &869565 &3245.91 &6557.38 & 7380.07\\ \hline
\multicolumn{6}{l}{Signal Region 2 Leptons (SR2L)}\\ \hline
1 Muon + 1 Electron &117.77 & 0.85& 0.85&0.05  & 0&0  \\ 
Charge =0&117.77 &0.43 & 0.85& 0.04 & 0& 0 \\
$P_{T}^{Muon} >4.0$GeV &101.66 &0.43 &0.85 & 0.03& 0  & 0 \\ 
$N_{trk}$ in $\Delta R_{lep-trk}>0.1$=0&101.66 &0.43 &0.85 & 0.03& 0  & 0 \\ 
 \hline
    \end{tabular}
    \caption{Event yield after each cut at 2\rm{$nb^{-1}$} for each sample generated (2LSR only). \label{table:2lep}}
  
    \end{center}
    \end{table*}

%
%
%
%

\end{document}